\documentclass[pra, floatfix, twocolumn]{revtex4-2}
\usepackage{graphicx}
\usepackage{color}
\usepackage{float}
\usepackage{amsmath, amsfonts, amssymb, bm, dsfont, mathtools}
\usepackage[bb=libus]{mathalpha}
\usepackage{hyperref}
\hypersetup{
	colorlinks=true,
	linkcolor=blue,
	citecolor=red,
	filecolor=magenta,
	urlcolor=blue,
}

\begin{document}
%%%%%%%%%%%%%%%%%%%%%%%%%%%%%%%%%%%%%%%%%%%%%%%%%%%%%%%%%%%%%%%%%
\title{Dynamically assisted Schwinger pair production in differently polarized\\ electric fields with the frequency chirping}
\author{Abhinav Jangir}
\email{2022rpy9087@mnit.ac.in}
\address{Department of Physics, MNIT Jaipur, Jaipur, Rajasthan, India}
\author{Anees Ahmed}
\email{anees.phy@mnit.ac.in}
\address{Department of Physics, MNIT Jaipur, Jaipur, Rajasthan, India}
\date{\today}
%%%%%%%%%%%%%%%%%%%%%%%%%%%%%%%%%%%%%%%%%%%%%%%%%%%%%%%%%%%%%%%%%%%
\begin{abstract}
	We investigate the enhanced dynamically assisted electron-positron pair production in differently polarized electric fields with frequency chirps within the real-time Dirac-Heisenberg-Wigner formalism. The combined influence of the chirp strength and the field polarization on the momentum distribution and the total number density of the created pairs is studied in detail for one-color fields as well as dynamically assisted two-color combined fields. Frequency chirps, applied in different field configurations, strongly reshape the momentum spectra by modifying the interference patterns and enhancing the peak momentum distribution. In the dynamically assisted case, the total number density can be enhanced significantly over $2-3$ orders when large chirps are applied to both strong and weak fields. Furthermore, we observe that sensitivity of the number density to field polarization progressively diminishes as the chirp strength increases, a trend that holds for both one-color field and the assisted two-color combined fields. A comparison of different chirping scenarios shows that chirping only the weak field produces nearly the same enhancement in the pair yield as chirping both fields simultaneously, demonstrating that the weak-field chirp plays the dominant role in the chirp-induced enhancement of the dynamically assisted Schwinger mechanism. We also find the enhancement factor to be the largest near circular polarization in the chirp-free and weakly chirped regimes, but is strongly suppressed with increasing weak-field chirp despite the continued growth of the absolute pair yield. These results provide new insight into the interplay among chirp, polarization, and dynamical assistance in strong-field quantum electrodynamics.
\end{abstract}
%%%%%%%%%%%%%%%%%%%%%%%%%%%%%%%%%%%%%%%%%%%%%%%%%%%%%%%%%%%%%%%%%%%
\maketitle
\section{Introduction}\label{sec:Intro}
	One of the novel features of Quantum Electrodynamics (QED) is its prediction of the creation of matter from light in the quantum vacuum, such as electron–positron ($e^+e^-$) pair production under ultra-strong electromagnetic fields. In this nonperturbative phenomenon, the vacuum behaves as a nonlinear medium that can decay into real particle-antiparticle pairs when subjected to sufficiently intense fields. This was first proposed by Sauter in 1931 \cite{Sauter1931Uber}, and studied in 1936 by Euler and Heisenberg \cite{heisenberg1936folgerungen}. Later in 1951, Schwinger extended the problem by employing the proper-time technique and calculated the pair production rate $\Gamma \sim \text{exp}(-\pi E_\text{cr}/E)$ for a constant electric field \cite{schwinger1951on}, where $E_\text{cr} = m^2c^3/e\hbar\approx 1.3 \times 10^{16}$ V/cm is Schwinger critical field strength ($m$ is the electron mass and $-e$ is the electron charge) above which the pair production rate becomes appreciable. However, due to the extremely high critical strength (corresponding to a laser intensity of $I_\text{cr} \sim 4.3 \times 10^{29}$ W/cm$^2$), researchers have not been able to observe this phenomena experimentally. Consequently, vacuum pair production in ultra-intense external fields, commonly referred to as the Schwinger effect, remains a major challenge for researchers due to its nonperturbative tunneling nature and the extraordinarily large field strengths required for direct observation. 
	
	The recent advances in ultra-intense laser technology \cite{marklund2009quantum, pike2014photon} are directed toward the experimental observation of $e^+e^-$ pair production. The application of chirped pulse amplification technique has significantly enhanced achievable laser intensities. For instance, present-day laser systems have reached intensities on the order of $10^{22}$ $\text{W/cm}^2$ \cite{yanovsky2008ultra}, while several planned facilities aim to achieve intensities in the range of $10^{25}$–$10^{26}$ $\text{W/cm}^2$ \cite{ringwald2001pair}. With the advent of next-generation high-intensity and ultrashort laser facilities, such as the Extreme Light Infrastructure (ELI) \cite{ELI} and the X-ray Free Electron Laser (XFEL) \cite{XFEL}, subcritical field strengths approaching $E \approx 0.1E_{\text{cr}}$ may become accessible. Consequently, experimental investigations of QED vacuum decay into $e^+e^-$ pairs could become feasible in the near future. See Refs. \cite{piazza2012extremely, fedotov2023advances} for the recent investigations on high-energy processes within the realm of relativistic quantum dynamics, QED, and nuclear and particle physics, occurring in extremely intense laser fields.

	In the last two decades, several methods have been developed to investigate the pair production from the vacuum instability in presence of the ultra-intense external fields. Notable methods include worldline instanton technique \cite{gies2005pair, dunne2006worldline, schneider2016dynamically, dumlu2011complex}, effective Lagrangian techniques \cite{dunne2005heisenberg}, Furry-picture quantization \cite{Aleksandrov2020pair, aleksandrov2017momentum}, quantum kinetic methods such as solving quantum Vlasov equation (QVE) \cite{kluger1998quantum, kim2011nonadiabatic, aleksandrov2024kinetic, huet2014vlasov, schmidt1998quantum, alkofer2001pair, nuriman2012enhanced, abdukerim2013effects} or its equivalent formulation in terms of the Riccati-type equation \cite{dumlu2010schwinger, marinov1977electron, dumlu2009quantum}, computational quantum field theory (CQFT) \cite{braun1999numerical, krekora2004klein, tang2013electron, li2021study}, S-matrix theory \cite{muller2003differential, muller2003nonlinear, deneke2008bound, he2012electron},  and the Dirac-Heisenberg-Wigner (DHW) formalism \cite{blinne2014pair, Li2015effects, li2015nonperturbative}, among others.
	
	Various field configurations with temporal and spatial pulse shaping have been the subject of extensive research. Hebenstreit \textit{et al.} discovered the extreme sensitivity of the longitudinal momentum spectra to the field parameters and its oscillations can be used as a probe of subcycle structure in ultrashort laser pulses \cite{Hebenstreit2009momentum}. Numerous studies have shown that both the dynamics of created particles and the electron-positron pair production rate are highly sensitive to the characteristics of the external field, including the pulse duration, carrier-envelope phase, number of pulses, and pulse shape (or envelope) \cite{aleksandrov2017pulse, abdukerim2013effects, Akkermans2012Ramsey, jangir2026carrier, chen2024asymmetric}. In addition, temporally oscillating and spatially inhomogeneous electric fields have been widely explored using a variety of theoretical approaches, revealing a  nontrivial dependence of pair production on the underlying field parameters \cite{kohlfurst2018pondermotive, Hebenstreit2011particle, kohlfurst2016effect, Aleksandrov2020pair, aleksandrov2016electron, ababekri2019effects, lv2018role, ababekri2020chirp, li2021study, li2021enhanced, bake2025vacuum}.

	Since an appreciable pair production rate in Schwinger effect requires extremely intense fields that have not yet been achieved experimentally, several catalytic mechanisms have been proposed to overcome this limitation and enable observable pair production under subcritical field strengths. Sch$\ddot{\text{u}}$tzhold \textit{et al.} \cite{schutzhold2008dynamically} proposed one such mechanism, known as the dynamically assisted Schwinger effect (DASE), which effectively combines a low-frequency strong field with a high-frequency weak field. This combined field configuration enhances the pair production rate by several orders of magnitude compared to a single-field setup. It has also been observed that the enhancement under DASE mechanism strongly depends on the shape of the fast pulse \cite{linder2015pulse}. DASE has been the focus of significant research efforts in the literature, where its combined effects have been analyzed for a wide range of field parameters for both temporally and spatially inhomogeneous field configurations. Representative works can be found in Refs. \cite{Orthaber2011momentum, li2021enhanced, li2021study, Olugh2020dynamically, nuriman2012enhanced, Bra2025RelativephaseDO}.
	
	Another method to enhance the pair production is the introduction of a frequency chirp in the electric field. For appropriate chirp parameters, the pair number density can be amplified by several orders of magnitude, and the resulting momentum spectra show pronounced sensitivity to the chirp.  Frequency-chirped electric fields have been investigated in spatially homogeneous but time-dependent field configurations \cite{dumlu2010schwinger, Olugh2019pair, Chen2025spin, Olugh2026Frequency} as well as in spatially inhomogeneous oscillating electric fields \cite{ababekri2020chirp, li2021enhanced}.   
	
	$e^+e^-$ pair production in a general elliptically polarized electric field has emerged as an important direction in strong-field QED research. Elliptically polarized fields are widely employed to probe distinctive features of strong-field ionization (SFI) dynamics that remain inaccessible in purely linearly polarized configurations \cite{hofmann2013comparison, hofmann2014interpreting}. Due to similarity between strong-field ionization and vacuum pair production, investigating pair creation under elliptical polarization not only reveals novel aspects of nonperturbative QED processes but also offers valuable comparative insight into SFI phenomena. While numerous theoretical works have analyzed the influence of laser-field polarization on pair production, experimental techniques have already achieved polarization degrees as high as $\pm 0.93$ in low-intensity electric fields \cite{pfeiffer2012attoclock}. In addition, several studies \cite{Li2015effects, Olugh2019pair, wang2021effect, Olugh2020asymmetric, Olugh2020dynamically, Olugh2026Frequency} have explored the interplay between polarization and other field parameters such as frequency chirp, temporal asymmetry, or the dynamically assisted Schwinger effect, demonstrating their combined impact on momentum spectra and particle yield.
	
	The aforementioned studies have established that frequency chirping, field polarization, and dynamical assistance each play an important role in determining the pair-production rate and the resulting momentum distributions. In particular, Olugh \textit{et al.} \cite{Olugh2019pair} investigated frequency-chirped elliptically polarized one-color electric fields within the DHW formalism and demonstrated that chirping and polarization jointly influence the momentum spectra and particle yield. However, frequency chirping in dynamically assisted two-color fields with different polarization configurations remains largely unexplored. In particular, the independent chirping of the strong low-frequency field and the weak high-frequency assisting field has not been systematically studied. Such a setup allows the individual roles of the two chirps in shaping interference patterns, momentum distributions, and pair-production enhancement to be disentangled.

	In this paper, we adapt the DHW formalism to study the $e^+e^-$ pair production in frequency-chirped dynamically assisted electric fields with various polarization configurations. We present numerical results for both the momentum distributions and the number densities of the produced particles in presence of either strong or weak one-color field with different chirp values. In addition, we analyze the dynamically assisted two-color combined field and discuss the corresponding numerical results. In particular, we address the following key questions : How does the momentum distributions evolve when a chirp is applied to the strong field only, weak field only, or both simultaneously in case of two-color combined field? Does chirping the weak field influence the pair yield differently from chirping the strong field? Furthermore, how does the polarization interplay with the strong or weak field chirp parameters in shaping the momentum distributions and total number density? By independently controlling the chirp of each field component, we disentangle their respective roles in modifying interference patterns, spectral shifts, and enhancement mechanisms in the dynamically assisted regime. The results provide a detailed characterization of how chirping and polarization jointly affect pair creation in dynamically assisted electric fields.
	
	The paper is organized as follows: In Sec. \ref{sec:Theory}, we introduce the electric field model, and briefly discuss the DHW formalism as our numerical approach. In Sec. \ref{sec:results_chirp_free}, we present our numerical results for the chirp free fields. Section \ref{sec:results_chirp_strong} is devoted to the case where chirping is applied to the strong field only, while Sec. \ref{sec:results_chirp_weak} discusses chirping in the weak field only. In Sec. \ref{sec:results_chirp_strong_weak}, we investigate the case where chirping is applied to both strong and weak fields, simultaneously. Finally, we summarize our findings and conclude in Sec. \ref{sec:Conclusion}.
\section{Background field and Theoretical formalism}\label{sec:Theory}
	\subsection{Model of the background electric field}\label{subsec:field_model}
	We consider the spatially homogeneous, time-dependent two-color dynamically assisted laser pulse with frequency chirps and arbitrary polarization as our model for the electric field. This is constructed as the combination of a strong-slowly varying field $\mathbf{E}_{1s}(t)$ with a weak-rapidly varying field $\mathbf{E}_{2w}(t)$, both sharing the same polarization state, as follows
	\begin{equation}\label{eq:field_model}
		\begin{split}
		\mathbf{E}(t) &= \mathbf{E}_{1s}(t) + \mathbf{E}_{2w}(t)\\
		& = \dfrac{E_{1s0}}{\sqrt{1 + \delta^2}}\,\text{exp}\left( -\frac{t^2}{2\tau^2} \right)\begin{pmatrix}
		\cos(b_1 t^2 + \omega_1 t) \\
		\delta \sin(b_1 t^2 + \omega_1 t)\\
		0
		\end{pmatrix}\\
	 	& \quad+ \dfrac{E_{2w0}}{\sqrt{1 + \delta^2}}\,\text{exp}\left( -\frac{t^2}{2\tau^2} \right)\begin{pmatrix}
		\cos(b_2 t^2 + \omega_2 t) \\
		\delta \sin(b_2 t^2 + \omega_2 t)\\
		0
		\end{pmatrix},		
		\end{split}
	\end{equation}
	where $E_{1s0}/\sqrt{1 + \delta^2}$ and $E_{2w0}/\sqrt{1 + \delta^2}$ denote the maximal field strength of the strong-slowly varying and weak-rapidly varying electric fields, respectively. To ensure the meaningful comparison across different polarizations, the peak field amplitude is normalized by the factor $1/\sqrt{1 + \delta^2}$. The pulse duration is $\tau$, while $\omega_{1,2}$ and $b_{1,2}$ are the field frequencies and frequency chirps of the strong and weak field, respectively. In the standard approach to DASE \cite{schutzhold2008dynamically}, we set $\omega_2 = k\omega_1$, where $k$ is the high harmonics order of the weak field relative to the strong field. In this work, we fix k = 7, placing the weak field in the high-frequency regime where dynamical assistance is effective, while maintaining computational feasibility and remaining consistent with values commonly used in previous studies, see the references listed in Sec. \ref{sec:Intro}.
	
\begin{figure*}[tbh]
		\centering
		\includegraphics[width=0.7\textwidth]{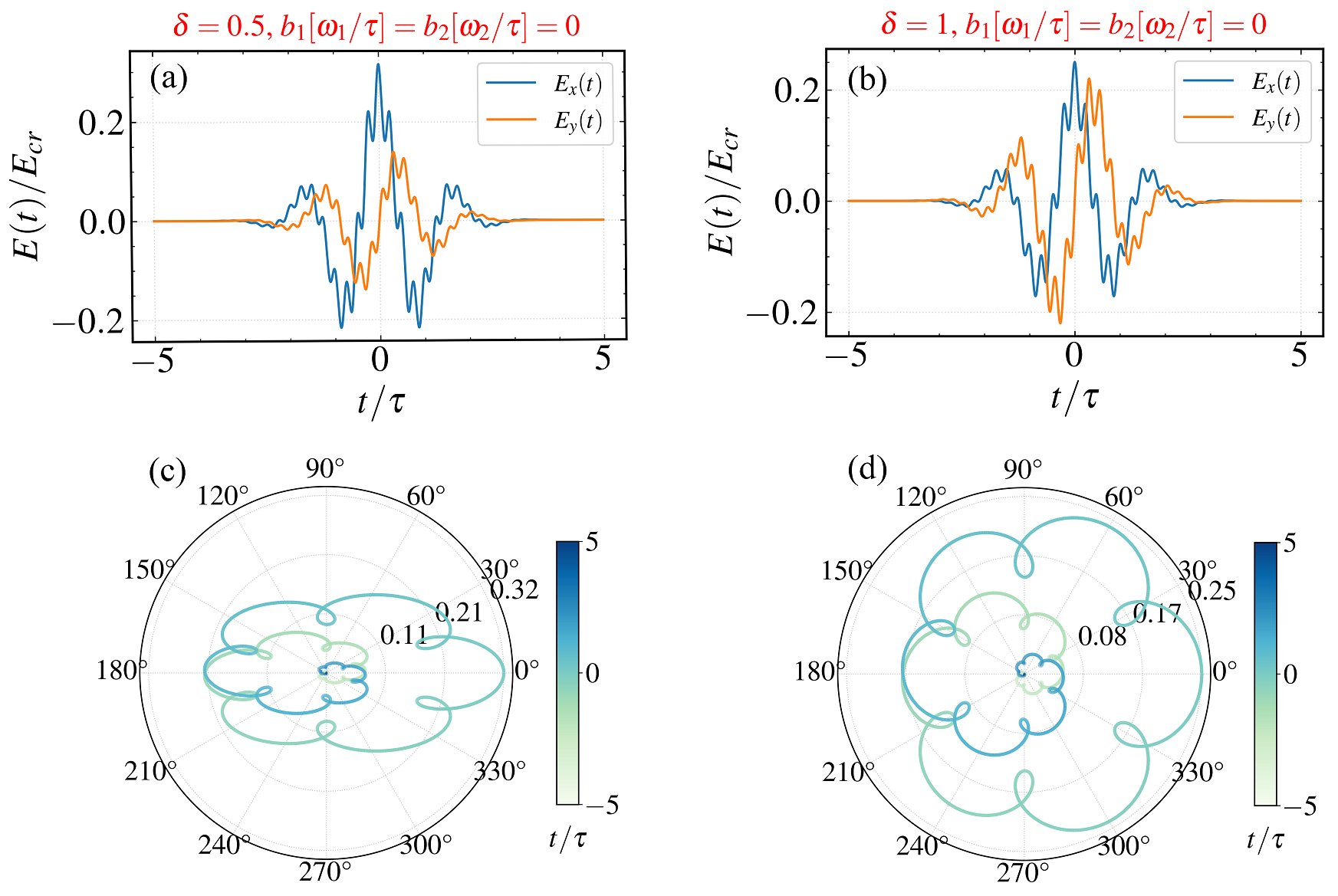}
		\caption{Electric field configuration described by Eq. \eqref{eq:field_model}. The top panels (a) and (b) show the time-dependence of the field components $E_x(t)$ and $E_y(t)$ for different polarization in chirp-free case ($b_1 = b_2 = 0$), while the bottom panels (c) and (d) present the corresponding polar diagrams of the field $\mathbf{E}(t)$. The  field parameters are same as given in Eq. \eqref{eq:parameters}.}
		\label{fig:field_chirp_free}
	\end{figure*}

\begin{figure*}[tbh]
		\centering
		\includegraphics[width=0.7\textwidth]{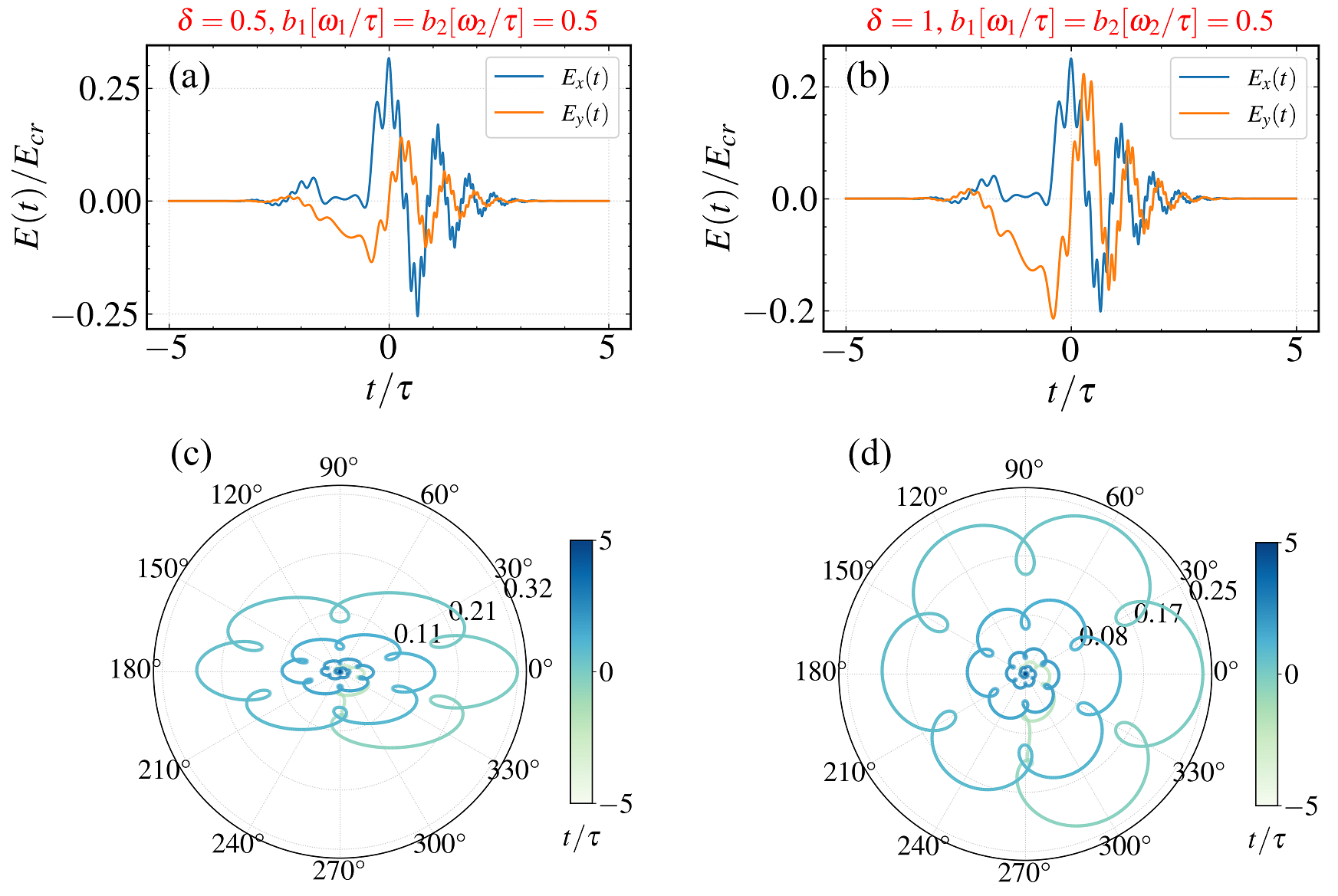}
		\caption{Same as Fig. \ref{fig:field_chirp_free} except for $b_1[\omega_1/\tau] = b_2[\omega_2/\tau] = 0.5$.}
		\label{fig:field_chirp}
	\end{figure*}

	The field polarization (or ellipticity) $\delta$ satisfies $|\delta| \leq 1$, with $\delta = 0$ corresponds to linear and $\delta = \pm1$ to circular polarization. The time evolution of the electric field \eqref{eq:field_model} is shown in Figs. \ref{fig:field_chirp_free} and \ref{fig:field_chirp} for fields without chirp and fields with chirp, respectively. 

	Throughout this paper, natural units ($\hbar = c = 1$) are used. Furthermore, in our numerical calculations, a set of parameters characterizing the electric field  \eqref{eq:field_model} are fixed as 
	\begin{equation}\label{eq:parameters}
		\begin{split}
			E_{1s0} &= 0.2\sqrt{2}\,E_\text{cr},\quad E_{2w0} = 0.25E_{1s0} = 0.05\sqrt{2}\,E_\text{cr},\\
			\omega_1 &= 0.09\,m,\quad \omega_2 = 7\omega_1 = 0.63\,m,\quad \tau = 40/m.\\
		\end{split}
	\end{equation}
	
	In this work, we assume identical pulse durations, $\tau_1 = \tau_2 = \tau$, a common simplification adopted in many previous studies of the dynamically assisted Schwinger effect. This choice reduces the dimensionality of the parameter space, allowing us to focus on the combined effects of field polarization and frequency chirp without introducing an additional independent parameter associated with the pulse durations. Allowing $\tau_1$ and $\tau_2$ to vary independently would modify the chirp parameters $b_i = a\omega_i/\tau_i$, as well as the effective Keldysh parameters and temporal overlap of the individual field components. Consequently, both the pair yield and the enhancement factor may be quantitatively affected. A systematic investigation of the combined influence of independently varying pulse durations, frequency chirp, and field polarization is beyond the scope of the present work and is left for future study.
	
	For subcritical fields $(E< E_\text{cr})$, the density of the created pairs remains sufficiently small, allowing the induced backreaction to be neglected. Detailed study of the backreaction effects can be found in Refs.~\cite{alkofer2001pair, block1999pair, tanji2009dynamical, jiang2023backreaction}.

	The purely time-dependent electric field \eqref{eq:field_model} should be considered as an idealized theoretical model that enables us to isolate specific behaviors of the system. If interpreted as the \textit{dipole approximation} to a realistic space-time dependent two-color standing-wave field generated by counterpropagating laser pulses, the approximation is expected to be valid when the characteristic spatial scales of the laser field, such as the focal width and carrier wavelength, are much larger than the pair-formation length scale. For a detailed discussion on the limitations of the dipole approximation, we direct readers to Refs. \cite{lv2018role, aleksandrov2017momentum}. These studies demonstrated that the relative error associated with the dipole approximation can reach several orders of magnitude for rapidly oscillating fields. Moreover, even within the nominal range of validity of the approximation, deviations from the full space-time-dependent treatment may remain significant, reaching approximately 100\%. It is also well established that the finite spatial extent of the field and its gradients can modify the total pair yield, reshape momentum distributions, and alter interference patterns through self-bunching and ponderomotive effects \cite{Hebenstreit2011particle, kohlfurst2018pondermotive}. Therefore, the results presented in this work should be understood as corresponding to the spatially homogeneous limit, while a fully space-time-dependent treatment is left for future investigation.
	
	Pair production can occur via two different processes or/and mechanisms. The Keldysh parameter $\gamma = m\omega/eE$ \cite{keldysh2024ionization, brezin1970pair} describes the dominant mechanism for the pair production. One mechanism is the tunneling process, also called the Schwinger mechanism \cite{schwinger1951on, schutzhold2008dynamically}, which dominates for $\gamma \ll 1$ (low-frequency strong fields). In contrast, the other one is the perturbative multiphoton absorption mechanism \cite{brezin1970pair, burke1997positron} which dominates for $\gamma \gg 1$ (high-frequency weak fields). While the Schwinger effect has yet to be experimentally observed, perturbative multiphoton absorption was first observed experimentally in 1997 at SLAC through the collision of a high-energy electron beam with an intense laser pulse \cite{burke1997positron, bamber1999studies}. For the intermediate regime $\gamma \sim \mathcal{O}(1)$, the multiphoton absorption becomes nonperturbative in nature and is known as the nonperturbative multiphoton regime \cite{Kohlfurst2014Effective, mocken2010nonperturbative}. 
	
	With the field parameters in Eq. \eqref{eq:parameters}, we obtain, for the strong-slowly varying field, the Keldysh parameter $\gamma_s = 0.32\sqrt{1 + \delta^2}$, and for weak-rapidly varying field, $\gamma_w = 8.91\sqrt{1 + \delta^2}$ for each polarization without chirp. These values indicate that pair production in the strong field predominantly occurs in the tunneling regime, whereas in the weak field it is primarily associated with multiphoton processes. For the two-color field $\mathbf{E}(t)$, the combined Keldysh parameter \cite{schutzhold2008dynamically} is $\gamma_c = \dfrac{m\omega_2}{eE_{1s0}}\sqrt{1 + \delta^2} = 2.23\sqrt{1 + \delta^2}$. 
	
	For nonzero chirps, the effective Keldysh parameter is calculated at a given time using the effective frequency $\omega_\text{eff} = \omega + bt$, where $b$ denotes the linear chirp parameter. We restrict our study to the \textit{normal chirp} regime \cite{Olugh2019pair, li2021enhanced, ababekri2020chirp}, for which the chirp parameter is defined as $b = a\omega/\tau$ with ${0 < a < 1}$. Without loss of generality, and in line with our exploratory objective of understanding qualitative analysis into the combined effects of chirp and polarization on pair production, we extend our analysis to the range $0 \leq a \leq 1$. This extension is intended to provide insight into the qualitative trends associated with increasingly strong frequency modulation within the DHW framework, rather than to establish quantitatively realistic predictions for experimentally relevant laser pulses. While the corresponding maximum chirp values, $b_1^\text{max} = \omega_1/\tau = 0.00225\,m^2$ for $\mathbf{E}_{1s}$, and $b_2^\text{max} = \omega_2/\tau = 0.01575\,m^2$ for $\mathbf{E}_{2w}$, correspond to the limiting case $a=1$, which has also been considered in previous studies \cite{Olugh2019pair}, the results obtained for these largest chirp values should therefore be interpreted as illustrating qualitative trends induced by strong chirping, rather than as quantitative predictions.

	\subsection{Theoretical description : The DHW formalism} \label{subsec:DHW_formalism}
	Our work is based on the real-time Dirac-Heisenberg-Wigner formalism that has been extensively used to investigate vacuum pair production in strong external background fields \cite{Birula1991phase, Hebenstreit2010Schwinger, Hebenstreit2011particle}. This method has the advantage of not only providing complete phase-space information for particle production in vacuum, but also being applicable to arbitrary background fields. The details of the derivation can be found in \cite{Kohlfuerst2015phd, Hebenstreit2011phd, Birula1991phase}; here, we summarize only the essential key steps relevant for the numerical calculations. We start by introducing the gauge-covariant density operator of Dirac field operator $\Psi$ in the Heisenberg picture
	\begin{equation}\label{eq:density operator}
		\hat{\mathcal{C}}_{\alpha\beta} = \mathcal{U}(A,r,s)[\bar{\Psi}_\beta(r-s/2), {\Psi}_\alpha(r+s/2)],
	\end{equation}
	where $r$ and $s$ are the center of mass and relative coordinates, respectively. The Wilson line factor
	\begin{equation}
		\mathcal{U}(A; r, s) = \exp \left[ ies \int_{-1/2}^{1/2} d\xi\, A(r + \xi s) \right].
	\end{equation}
	is introduced (which is related to the elementary charge $e$ and the background gauge field $A$) to ensure the gauge invariance of the density operator. The heart of the DHW approach is the quantity known as the covariant Wigner operator, which is defined as the Fourier transform of the density operator Eq. \eqref{eq:density operator} with respect to the relative coordinate $s$ 
	\begin{equation}\label{eq:covariant Wigner operaotr}
		\widehat{\mathcal{W}}_{\alpha\beta}(r, p) = \frac{1}{2} \int d^4s\, e^{ip s} \widehat{\mathcal{C}}_{\alpha\beta}(r, s).
	\end{equation}
	Since the focus is on the vacuum pair creation, taking the vacuum expectation value of the Eq. \eqref{eq:covariant Wigner operaotr} allows us to switch from the covariant Wigner operator  $\widehat{\mathcal{W}}(r, p)$ to the covariant Wigner function (or simply the Wigner function) $\mathbb{W}(r, p)$
	\begin{equation}\label{eq:covariant Wigner function}
		\mathbb{W}(r, p) = \langle \Phi | \widehat{\mathcal{W}}(r, p) | \Phi \rangle.
	\end{equation}
	The Winger function follows the Dirac algebra and can be decomposed in terms of the the Dirac bilinears. Thus, taking a spin decomposition on the Wigner function yields 16 covariant Wigner coefficients.
	\begin{equation}\label{eq:Wigner function components}
		\mathbb{W} = \frac{1}{4} \left( \mathds{1} \mathbb{S} + i\gamma_5 \mathbb{P} + \gamma^\mu \mathbb{V}_\mu + \gamma^\mu \gamma_5 	\mathbb{A}_\mu + \sigma_{\mu\nu} \mathbb{T}^{\mu\nu} \right).
	\end{equation}
	Under orthochronous Lorentz transformations, these coefficients transform as a scalar $\mathbb{S}$, pseudoscalar $\mathbb{P}$, vector $\mathbb{V}_\mu$, axial vector $\mathbb{A}_\mu$ and tensor $\mathbb{T}_{\mu\nu}$. They are related to the physical quantities as follows: $\mathbb{S}$ corresponds to the mass density, $\mathbb{P}$ to the pseudoscalar condensate density, $\mathbb{V}_\mu$ to the net fermion current density, $\mathbb{A}_\mu$ to the polarization density, and $\mathbb{T}_{\mu\nu}$ is related to the electric dipole-moment density. 
	The Wigner function follows the equations of motion \cite{Hebenstreit2011phd, Kohlfuerst2015phd}, 
	\begin{equation}\label{eq:dynamical EOM of W}
		D_t \mathbb{W} = -\frac{1}{2} \textbf{D}_x [\gamma^0 \bm{\gamma}, \mathbb{W}] + i m [\gamma^0, \mathbb{W}] - i \textbf{P} \{ \gamma^0 	\bm{\gamma}, \mathbb{W} \},
	\end{equation}
	where $D_t$, $\textbf{D}_x$, and $\textbf{P}$ represent the pseudodifferential operators.
	
	By substituting Eq. \eqref{eq:Wigner function components} into Eq. \eqref{eq:dynamical EOM of W}, we can obtain a set of partial differential equations for the 16 Wigner components. Since the field profile in Eq. \eqref{eq:field_model} is spatially homogeneous and time dependent; by applying the method of characteristics \cite{blinne2014pair, Blinne2016Comparison} and replacing the kinetic momentum $\textbf{p}$ with the canonical momentum $\textbf{k}$, reduces the 16 partial differential equations for the Wigner coefficients to 10 ordinary differential equations for the non-vanishing equal-time Wigner components, given by
	\begin{equation}
		\mathbb{w} = (\mathbb{s},\mathbb{v}_i,\mathbb{a}_i,\mathbb{t}_i),\quad \mathbb{t}_i:=\mathbb{t}_{0i} -\mathbb{t}_{i0}.
	\end{equation}
	Correspondingly, the non-vanishing vacuum initial conditions are 
	\begin{equation}
		\mathbb{s}_{\mathrm{vac}} = -\frac{2m}{\sqrt{\textbf{p}^2 + m^2}}, \qquad
		\mathbb{v}_{i,\mathrm{vac}} = -\frac{2p_i}{\sqrt{\textbf{p}^2 + m^2}}.	
	\end{equation}
	These 10 equations of motions are rather lengthy, we refer the readers to Refs. \cite{Hebenstreit2011phd, Kohlfuerst2015phd} for their detailed forms. 
	
	The single-particle momentum distribution function is defined as
	\begin{equation}\label{eq:distribution function}
		f_\textbf{k}(t) = \frac{1}{2\Omega_\textbf{k}(t)} \left( \varepsilon - \varepsilon_{\mathrm{vac}} \right),
	\end{equation}
	where $\Omega_\mathbf{k}(t) = \sqrt{\mathbf{p}^2 + m^2} = \sqrt{(\mathbf{k} - e\mathbf{A})^2 + m^2}$, is the total energy of the particles, where $\mathbf{A}(t)$ denotes the vector potential corresponding to the electric field $\mathbf{E}(t)$ of Eq. \eqref{eq:field_model}. $\varepsilon = m \mathbb{s} + p_i \mathbb{v}_i$ denotes the phase space energy density. Further, we define an auxiliary three dimensional vector 
	\begin{equation}
		\textbf{v}_\textbf{k}(t) \coloneqq \mathbb{v}_{i\textbf{k}}(t) - (1 - f_\textbf{k}(t))\mathbb{v}_{i\textbf{k},\text{vac}}(t). 
	\end{equation}
	The one-particle momentum distribution function $f_\textbf{k}(t)$ can then be obtained by solving the following ten ordinary differential equations (ODEs) for $f$ and the other nine auxiliary quantities $\textbf{v},\textbf{a}\coloneqq\mathbb{a}_i,\textbf{t}\coloneqq \mathbb{t}_i$:
	\begin{equation}\label{eq:DHW-10 ODEs}
	\begin{split}
		\dot{f} &= \frac{e\,\mathbf{E}\cdot\mathbf{v}}{2\,\Omega}, \\[6pt]
		\dot{\mathbf{v}} &= \frac{2}{\Omega^{3}}
		\Big[ (e\mathbf{E}\cdot\mathbf{p})\,\mathbf{p} - e\,\mathbf{E}\,\Omega^{2} \Big]\,(f-1)
		- \frac{(e\mathbf{E}\cdot\mathbf{v})\,\mathbf{p}}{\Omega^{2}}\\
		&\hspace{0.5cm} - 2\,\mathbf{p}\times\mathbf{a} - 2m\,\mathbf{t}, \\[6pt]
		\dot{\mathbf{a}} &= -\,2\,\mathbf{p}\times\mathbf{v}, \\[6pt]
		\dot{\mathbf{t}} &= \frac{2}{m}\Big[ m^{2}\mathbf{v} + (\mathbf{p}\cdot\mathbf{v})\,\mathbf{p} \Big],
	\end{split}
	\end{equation}
	with the initial conditions $f_\mathbf{k}(-\infty) = \mathbf{v}_\mathbf{k}(-\infty) = \mathbf{a}_\mathbf{k}(-\infty) = \mathbf{t}_\mathbf{k}(-\infty) = 0$, where an overdot represents the total time derivative. 
	
	The asymptotic momentum distribution of produced pairs is obtained in the limit $t\rightarrow + \infty$:
	\begin{equation}
		f_\mathbf{k}(\infty) = \lim_{t\rightarrow+\infty}f_\mathbf{k}(t).
	\end{equation}
	Finally, the total number of produced pairs per unit volume (in units of $m^3$) is calculated by integrating the asymptotic momentum distribution function $f_\mathbf{k}(\infty)$ over full momentum space:
	\begin{equation}
		N = \int \dfrac{d^3\mathbf{k}}{(2\pi)^3}f_\mathbf{k}(\infty).
	\end{equation}

	The coupled ODE system in Eq. \eqref{eq:DHW-10 ODEs} is solved numerically using Runge-Kutta algorithms of various orders, including both adaptive and fixed-step schemes, over the interval $t \in [-\eta\tau, \eta\tau]$, where $\eta>0$ is chosen sufficiently large to ensure that the external field vanishes at the integration boundaries. The numerical implementation was benchmarked by reproducing the momentum distributions reported in Ref.~\cite{Olugh2019pair}, yielding excellent agreement with the published results.
		
	The momentum integration for the total number density is performed using a momentum cutoff of $k_{\mathrm{cutoff}}=10\ m$ and a uniform momentum grid spacing of $\Delta k=0.08\ m$. The numerical convergence of the calculated number densities was verified by increasing the momentum cutoff to $k_{\mathrm{cutoff}}=15\ m$ and refining the momentum grid to $\Delta k=0.04\ m$ for representative cases, including those with the largest particle yields. In all cases examined, the resulting number densities differed by less than $0.3\%$, demonstrating that the adopted momentum discretization provides well-converged results. We further verified that the numerical solutions are insensitive to the choice of algorithm, the integration step size, and variations in $\eta$, confirming the stability and robustness of the numerical implementation.
	
\section{Numerical Results: Chirp-Free Fields }\label{sec:results_chirp_free}
	In this section, we present our numerical results for the one-color field configurations $\mathbf{E}_{1s}(t)$ and $\mathbf{E}_{2w}(t)$, as well as for the two-color combined field $\mathbf{E}(t)$. The effects of chirp are not considered in this section. 
	\subsection{One-color strong field $\mathbf{E}_{1s}(t)$}\label{subsec:results_chirp_free_strong_field}
\begin{figure}[t]
		\centering
		\includegraphics[width=\linewidth]{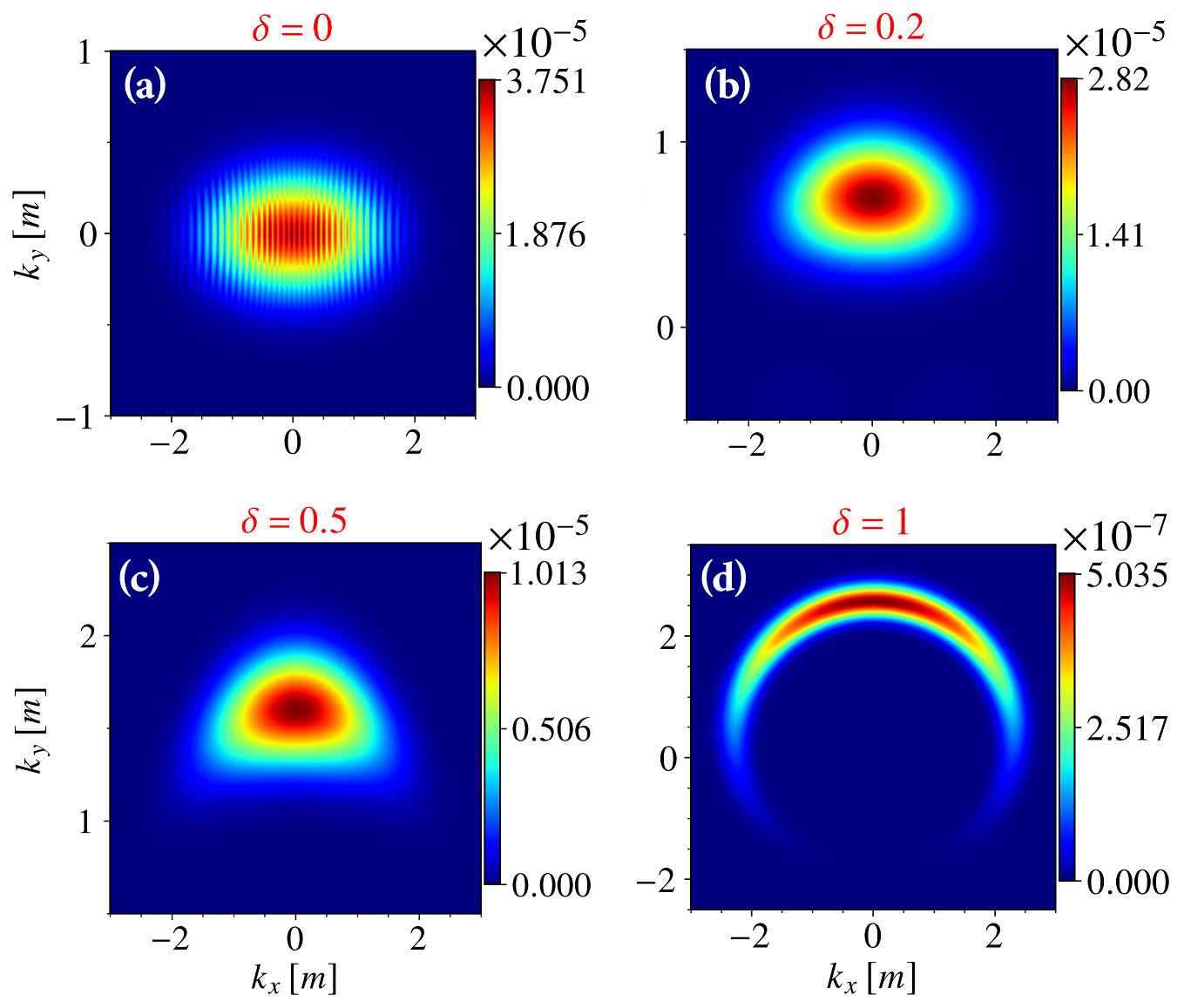}
		\caption{Momentum distribution of the created $e^+e^-$ pairs with $k_z = 0$ in the one-color strong field $\mathbf{E}_{1s}$. The plot shows the variation with $\delta$ in a chirp free case $(b_1 = 0)$. The other field parameters are same as Eq. \eqref{eq:parameters}.}
		\label{fig:MD_chirp_free_strong_field}
	\end{figure}

	In this subsection, we present our numerical results for the momentum distribution of the created $e^+e^-$ pairs in presence of the one-color strong field 
	\begin{equation}
		\mathbf{E}_{1s}(t)  = \dfrac{E_{1s0}}{\sqrt{1 + \delta^2}}\,\text{exp}\left( -\frac{t^2}{2\tau^2} \right)\begin{pmatrix}
			\cos(b_1 t^2 + \omega_1 t) \\
			\delta \sin(b_1 t^2 + \omega_1 t)\\
			0\end{pmatrix},
	\end{equation}
	in chirp-free case ($b_1 = 0$), with variation in $\delta$. With the field parameter values and corresponding $\gamma_s$ discussed in Subsec. \ref{subsec:field_model}, this field configuration lies in regime of Schwinger effect i.e., tunneling regime. The momentum distribution for different values of $\delta$ is shown in Fig. \ref{fig:MD_chirp_free_strong_field}. 

	In Fig. \ref{fig:MD_chirp_free_strong_field}(a), the field is polarized along the $x$-direction. The corresponding momentum distribution is symmetric with respect to $k_y = 0$, and is strongly but symmetrically elongated along the $k_x$ axis. The appearance of fine oscillatory structure in the distribution is a consequence of temporal interference between multiple pair-creation events during the pulse. This result is similar to the one in Ref. \cite{Hebenstreit2009momentum}. The distribution is overall centered about zero momentum, consistent with the absence of a net transverse force as the asymptotic value of the vector potential is chosen to be $A(\infty) = 0$ (which also removes the distinction between canonical and kinetic momentum at asymptotic times). 

	With the introduction of the transverse component in the field i.e., $\delta\neq0$, the momentum distribution undergoes a qualitative change. The symmetry along the $k_y=0$ axis is broken and the distribution shifts towards the finite $k_y$ region, indicating the onset of transverse drift induced by the rotating field. Simultaneously, the oscillatory interference pattern becomes suppressed, and with the increment in $\delta$, the distribution broadens in the momentum space as particles acquire momentum from both polarization components. The interference effect in the momentum distribution of produced pairs is related to the electric field profile. The variation of the field polarization can influence the dominant turning points $t_p$ obtained by solving the condition $\Omega_\mathbf{k}(t_p) = 0$ (see Refs. \cite{dumlu2011complex, dumlu2011interference} for analysis of interference effects in momentum spectra, based on turning points). The suppression of the interference pattern from Fig. \ref{fig:MD_chirp_free_strong_field}(a) to (d) is due to the departure of the dominating pairs of turning points from the real time $t$-axis when the field polarization is changed from linear to circular, and hence the interference effects between those pairs of turning points become weaker and even vanishing with increase in field polarization \cite{Olugh2020dynamically, Li2015effects}. 
	We also observe an overall decrease in peak momentum distribution value as $\delta$ increases. 
	
	In the circular polarization limit $\delta = 1$ (Fig. \ref{fig:MD_chirp_free_strong_field}(d)), the momentum distribution transforms into a crescent-like structure while the peak momentum distribution value is strongly suppressed compared to the linearly polarized case. The gradual decrease in the peak value with increase in the polarization $\delta$ is because of the fact that pair production, in tunneling regime, is exponentially sensitive to the instantaneous field strength; as the field becomes elliptical or circular, its direction rotates and the effective tunneling component along any fixed direction decreases, leading to exponential suppression and redistribution of probability over a broader ring structure, thereby lowering the maximum peak. A similar observation based on the variation of the polarization parameter for the one-color field, with parameters rendering the field in the tunneling regime, was observed in Ref. \cite{Li2015effects}. 
	
	\subsection{One-color weak field $\mathbf{E}_{2w}(t)$}\label{subsec:results_chirp_free_weak_field}
	In this subsection, we investigate the momentum distribution of the created $e^+e^-$ pairs in presence of the one-color weak field
		\begin{equation}
		\mathbf{E}_{2w}(t)  = \dfrac{E_{2w0}}{\sqrt{1 + \delta^2}}\,\text{exp}\left( -\frac{t^2}{2\tau^2} \right)\begin{pmatrix}
			\cos(b_2 t^2 + \omega_2 t) \\
			\delta \sin(b_2 t^2 + \omega_2 t)\\
			0\end{pmatrix},
	\end{equation}
	in chirp-free case $(b_2 = 0)$, with variation in $\delta$. 
	\begin{figure}[b]
		\centering
		\includegraphics[width=\linewidth]{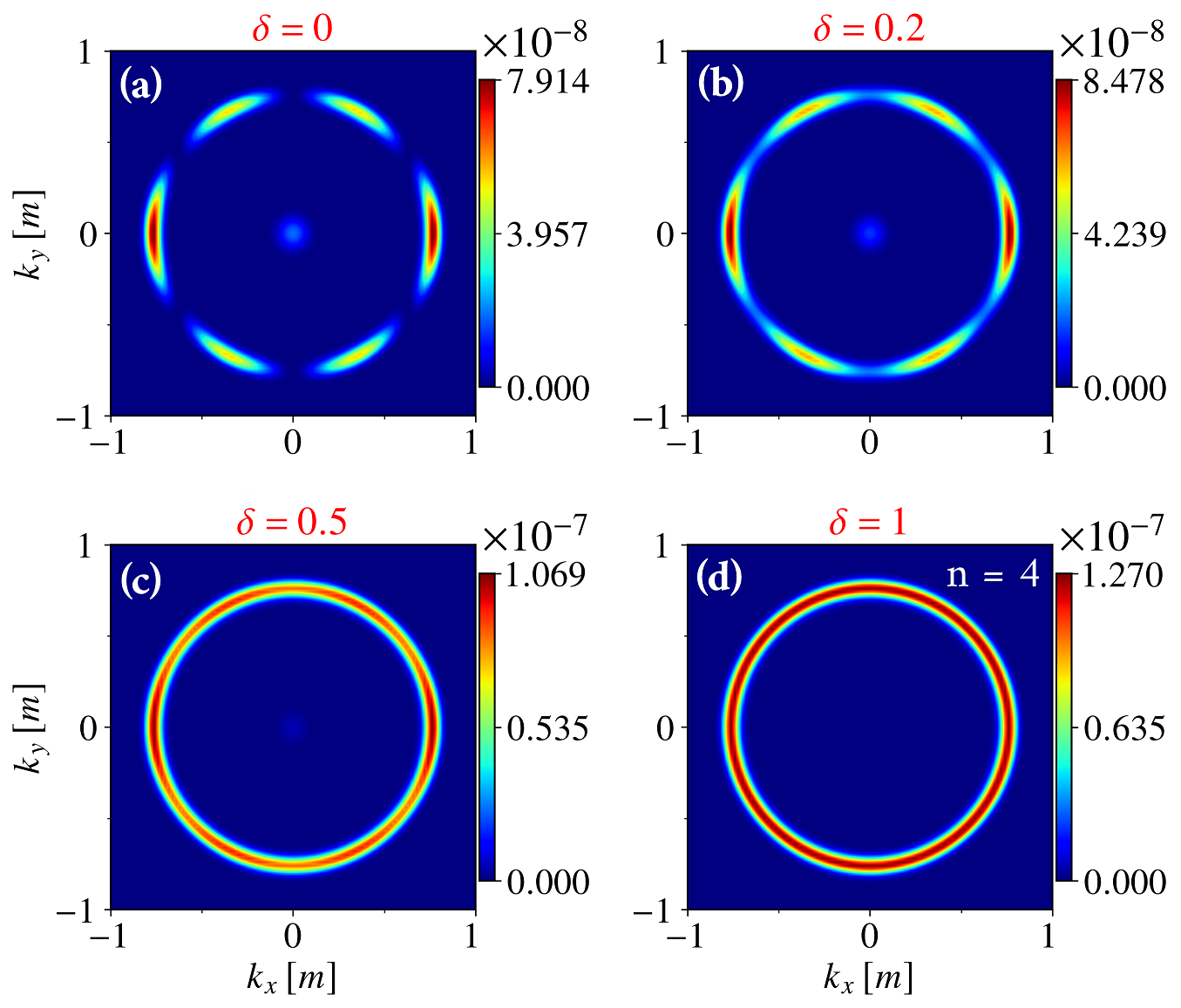}
		\caption{Momentum distribution of the created $e^+e^-$ pairs with $k_z = 0$ in the one-color weak field $\mathbf{E}_{2w}$. The plot shows the variation with $\delta$ in a chirp-free case $(b_2 = 0)$. The other field parameters are same as Eq. \eqref{eq:parameters}}.
		\label{fig:MD_chirp_free_weak_field}
	\end{figure}
	With the field parameter values and corresponding $\gamma_w$ discussed in Subsec. \ref{subsec:field_model}, the weak field configuration lies in the multiphoton absorption regime. This is also evident from the pronounced ring-like structure appearing in the momentum distribution shown in Fig. \ref{fig:MD_chirp_free_weak_field}, contrary to the strong field case in Fig. \ref{fig:MD_chirp_free_strong_field}.
	
	For small values of polarization i.e., $\delta=0$ and $0.2$, the distribution is split into multiple segments scattered symmetrically making a fragmented ring-like structure in the $(k_x,k_y)$ plane. As the value of $\delta$ is increased, these segments fuse together and the distribution is progressively smoothed out to evolve toward a more uniform ring as shown in Fig. \ref{fig:MD_chirp_free_weak_field}(d). This ring-like structure is formed due to the multiphoton absorption process \cite{Blinne2017Phd}. The radius of this ring, for linearly polarized field ($\delta = 0$), follows from energy conservation:
	\begin{equation}\label{eq:multiphoton_radius}
		|\mathbf{k}| = \sqrt{\left(\dfrac{n\omega}{2}\right)^2 - m^2_*},
	\end{equation}
	where $m_* = m\sqrt{1 + \dfrac{e^2E_0^2}{2m^2\omega^2}}$  is the effective mass and $n$ denotes the number of absorbed photons \cite{Kohlfurst2014Effective}. The energy conservation expression has a physical meaning that the absorption of $n$ photons of frequency $\omega$, shared between the electron and the positron, produces a characteristic excess momentum $\mathbf{k}$. See Refs. \cite{wollert2015spin, aleksandrov2018dynamically, Aleksandrov2020pair, mocken2010nonperturbative} for other applications of \eqref{eq:multiphoton_radius} and its generalizations. 
	
	With the laser intensity kept constant throughout this work, we nevertheless employ the effective-mass expression, although it is formally derived for monochromatic plane-wave backgrounds. Its purpose here is not to provide a rigorous description of the present polarized field configuration, but rather to serve as a simple reference for estimating the radius of the observed multiphoton ring. The corresponding effective mass is defined by
	\begin{equation}
		m_* = m\left[1 + \frac{m^2\langle -A_\mu A^\mu\rangle}{E_\mathrm{cr}^2}\right]^{1/2},
	\end{equation}
	where $A_\mu$ is the four-vector potential~\cite{Kohlfurst2014Effective}. For the field parameters given in Eq.~\eqref{eq:parameters}, the effective-mass model predicts the four-photon process ($n=4$), corresponding to a theoretical ring radius of $|\mathbf{k}|_{\mathrm{theoretical}}=0.7624\,m$, while the numerically extracted radius from Fig.~\ref{fig:MD_chirp_free_weak_field}(d) is $|\mathbf{k}|_{\mathrm{numerical}}=0.7611\,m$. A comparison between the theoretical and numerical radii of higher-order rings is presented in Table~\ref{tab:multiphoton_rings_table}.
	
	It should be emphasized that the effective-mass expression is derived under several idealized assumptions, including monochromatic plane-wave backgrounds and the long-pulse (effectively infinite pulse duration) limit, and it does not explicitly incorporate the polarization parameter used in the present field configuration. Consequently, the comparison should be regarded as a consistency check rather than a rigorous validation of the effective-mass model. The close agreement between the theoretical and numerical ring radii indicates that the effective-mass approximation nevertheless provides a useful estimate of the multiphoton ring radius in the present calculations. The remaining small discrepancy is naturally attributable to the idealized assumptions underlying the effective-mass model, rather than to the numerical calculations themselves.
	\begin{table}[tbh]
		\caption{Comparison between theoretical and numerical radii of multiphoton rings for the one-color weak fields $\mathbf{E}_{2w}$ obtained at $\delta=1$ without chirp $(b_2 = 0)$. The other field parameters are same as Eq. \eqref{eq:parameters}.}
		\label{tab:multiphoton_rings_table}
		\begin{ruledtabular}
			\begin{tabular}{ccc}
				$n$ & $|\mathbf{k}|_{\text{theoretical}}[m]$ & $|\mathbf{k}|_{\text{numerical}}[m]$ \\
				\hline
				4 & 0.7624 & 0.7611 \\
				5 & 1.2142 & 1.2120 \\
				6 & 1.6018 & 1.6001 \\
			\end{tabular}
		\end{ruledtabular}
	\end{table}
	
	The peak value of the asymptotic momentum distribution function $f_\mathbf{k}(\infty)$ exhibits a monotonic increment as the field polarization $\delta$ is increased. This result is contrary to the case of one-color strong field, see Fig. \ref{fig:MD_chirp_free_strong_field} where the peak value decreases with field polarization. The difference originates from the fact that the weak field operates in the multiphoton regime, rather than the tunneling regime. As the transverse component of the field is introduced ($\delta\neq0$), the electric field is rotated continuously, allowing particles to gain momentum in multiple directions during the pulse. This enhances the angular accessibility of multiphoton absorption channels. Consequently, the momentum distribution becomes more uniformly distributed along the resonant ring, leading to an increase in the peak value. 

	\subsection{Two-color combined field $\mathbf{E}(t)$}\label{subsec:results_chirp_free_combined_field}
	In this subsection, we investigate the $e^+e^-$ pair creation in presence of the dynamically assisted two-color combined field $\mathbf{E}(t) = \mathbf{E}_{1s}(t) + \mathbf{E}_{2w}(t)$, given by Eq. \eqref{eq:field_model} but in chirp-free case ($b_1 = b_2 = 0$). We present the numerical results for the momentum distribution and the number density of the created $e^+e^-$ pairs, as well as the enhancement factor, all with the variation in the field polarization $\delta$. 
	\begin{figure}[b]
		\centering
		\includegraphics[width=\linewidth]{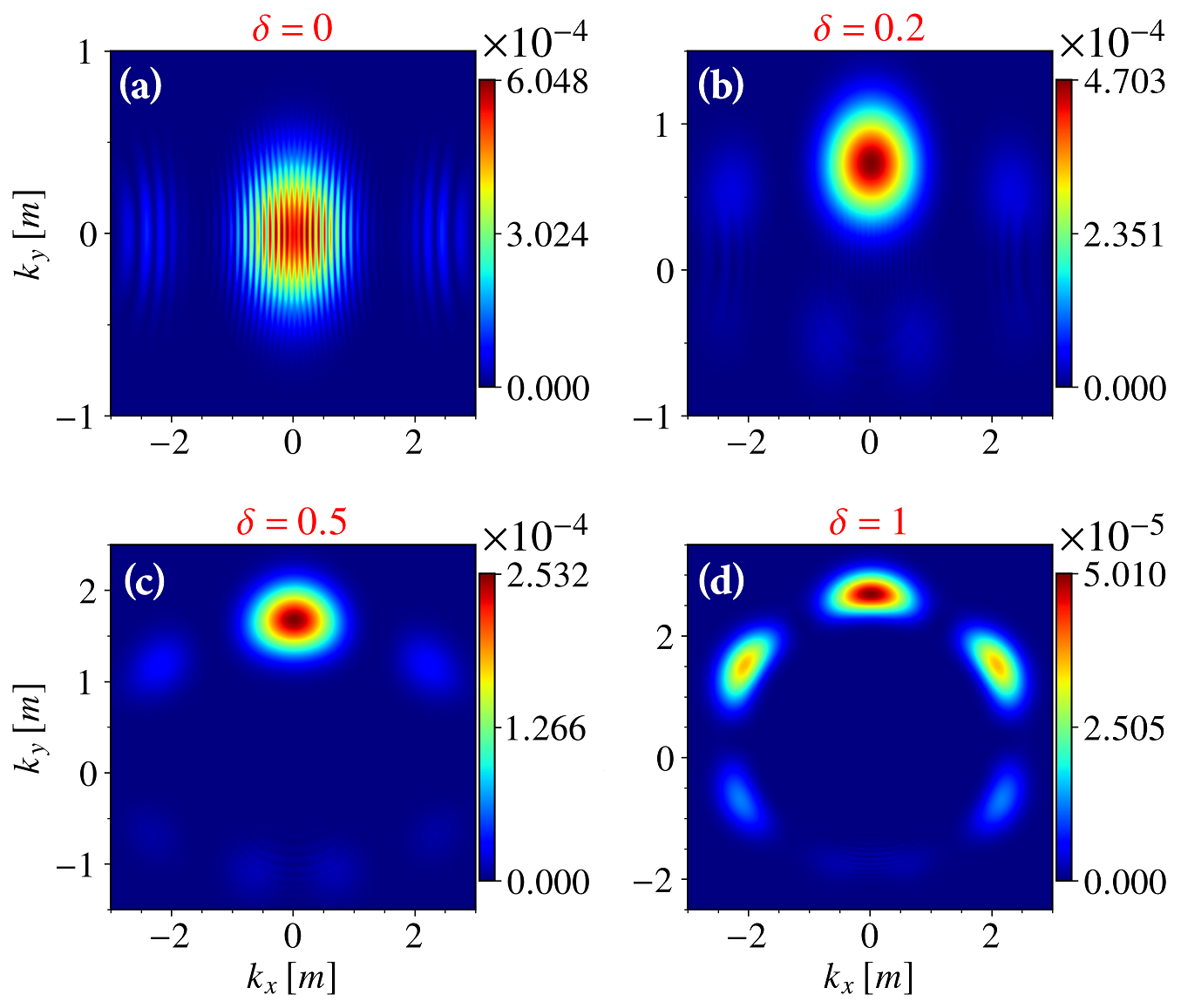}
		\caption{Momentum distribution of the created $e^+e^-$ pairs with $k_z = 0$ in the two-color combined field $\mathbf{E}(t)$. The plot shows the variation with $\delta$ in chirp-free fields ($b_1 = b_2 = 0$). The other field parameters are same as Eq. \eqref{eq:parameters}.}
		\label{fig:MD_chirp_free_combined_field}
	\end{figure}
	Fig. \ref{fig:MD_chirp_free_combined_field} shows the asymptotic momentum distribution function $f_\mathbf{k}(\infty)$ in the $(k_x,k_y)$ plane with variation in $\delta$.
	
	A very clear observation is the peak value of the momentum distribution function in overall figure is comparatively larger than one-color strong field or one-color weak field individually, which clearly states the dynamically assisted Schwinger mechanism \cite{Orthaber2011momentum, Olugh2020dynamically}. 
	
	For linear polarization ($\delta = 0$), the distribution is symmetric about $k_y =0$, and is elongated along the $k_x$ direction, with pronounced interference fringes, similar to the one-color strong field, cf. Fig. \ref{fig:MD_chirp_free_strong_field}(a). This indicates the efficient dynamic assistance, where the weak field lowers the effective tunneling barrier of the strong field and enhances pair production while preserving the interference structure of linearly polarized fields. As the ellipticity parameter is increased, the distributions exhibit a qualitative change: the central interference fringes are heavily suppressed, and the distribution shifts towards finite $k_y$ region, reflecting the growing influence of the rotating electric field and the resulting transverse drift of the produced particles. In the circular polarization limit $\delta = 1$, Fig. \ref{fig:MD_chirp_free_combined_field}(d), the distribution transforms like a fragmented ring-like structure with several localized intensity maxima. This is the indication of a competition between tunneling-dominated dynamics from the strong field and multiphoton absorption from the weak field. The suppression of the momentum distribution peak values in this limit demonstrates that, although dynamical assistance remains operative, the dynamics still retain significant nonperturbative character $(\gamma_c\sim\mathcal{O}(1))$. Hence, increasing the field polarization significantly weakens the tunneling efficiency and redistributes the pair yield over a broader region of momentum space. The multiple peaks observed at circular polarization, as shown in Figs. \ref{fig:MD_chirp_free_combined_field}(d), correspond to the so-called shell structures \cite{Otto2015lifting}, which have also been reported in \cite{Olugh2020dynamically}.

	To further analyze our study, we investigate the influence of the field polarization on the total number of produced particles in presence of the one-color weak field $\mathbf{E}_{2w}$, one-color strong field $\mathbf{E}_{1s}$ and two-color combined field $\mathbf{E} = \mathbf{E}_{1s} + \mathbf{E}_{2w}$, all in the chirp-free case. The corresponding number density behavior as a function of $\delta$ is shown in Fig. \ref{fig:ND,EF_chirp_free_combined_field}(a). We observe that the number density produced by the strong field dominates over the weak field contributions over all $\delta$, while the combined field yields a substantially larger number density than either field alone, clearly demonstrating the effectiveness of dynamical assistance. 
	
	\begin{figure}[tbh]
		\centering
		\includegraphics[width=\linewidth]{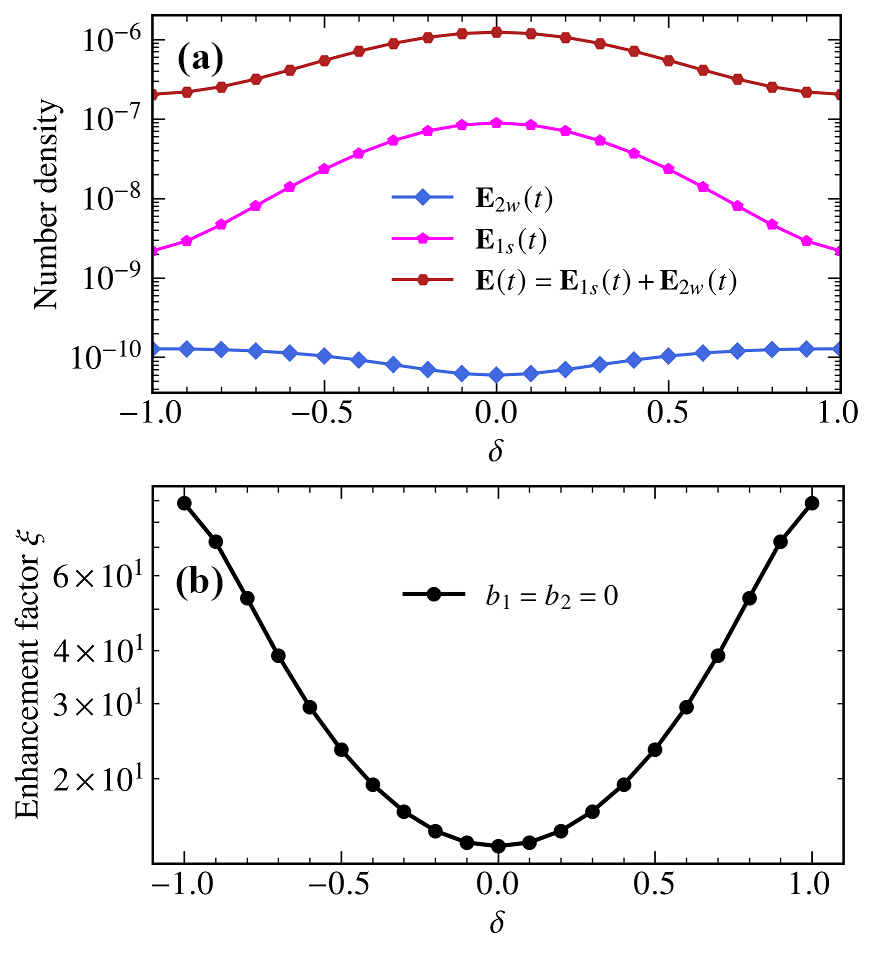}
		\caption{Chirp-free case ($b_1 = b_2 = 0$). (a) The number destiny (in units of $m^3$) of created particles as a function of field polarization $\delta$ for the weak field $\mathbf{E}_{2w}$, strong field $\mathbf{E}_{1s}$, and the combined field $\mathbf{E} = \mathbf{E}_{1s} + \mathbf{E}_{2w}$. (b) Corresponding enhancement factor of the total number as a function of $\delta$ in the combined field $\mathbf{E}$. The other field parameters are same as Eq. \eqref{eq:parameters}.}
		\label{fig:ND,EF_chirp_free_combined_field}
	\end{figure}
	
	In case of the weak field $\mathbf{E}_{2w}$, the number density is monotonically increased as $|\delta|$ is increased. This is consistent with the corresponding momentum distribution plots as shown in Fig. \ref{fig:MD_chirp_free_weak_field}. While in case of the strong field $\mathbf{E}_{1s}$ and the combined field $\mathbf{E}$, the behavior of the number density on the field polarization is opposite to the case of the weak field $\mathbf{E}_{2w}$, i.e., the number of produced particles in case of strong and combined field decrease as $|\delta|$ is increased. This behavior can be explained by the fact that in case of the strong field, the pairs are produced by the nonperturbative tunneling process, which is highly sensitive to the instantaneous field strength along a fixed direction. As $\delta$ is increased, the field vector is rotated, as shown in Fig. \ref{fig:field_chirp_free} and \ref{fig:field_chirp}, which reduces the time during which the field remains strong in any given direction. This effectively suppresses tunneling, leading to a monotonic decrease in the number density with increasing $\delta$. A similar behavior is observed in case of the combined field. 
	 
	We also observe that the sensitivity of the number density on the field polarization is strongest for the the strong field, weaker for the combined field, and weakest for the weak field. 
	\begin{figure*}[tbh]
		\centering
		\includegraphics[width=\linewidth]{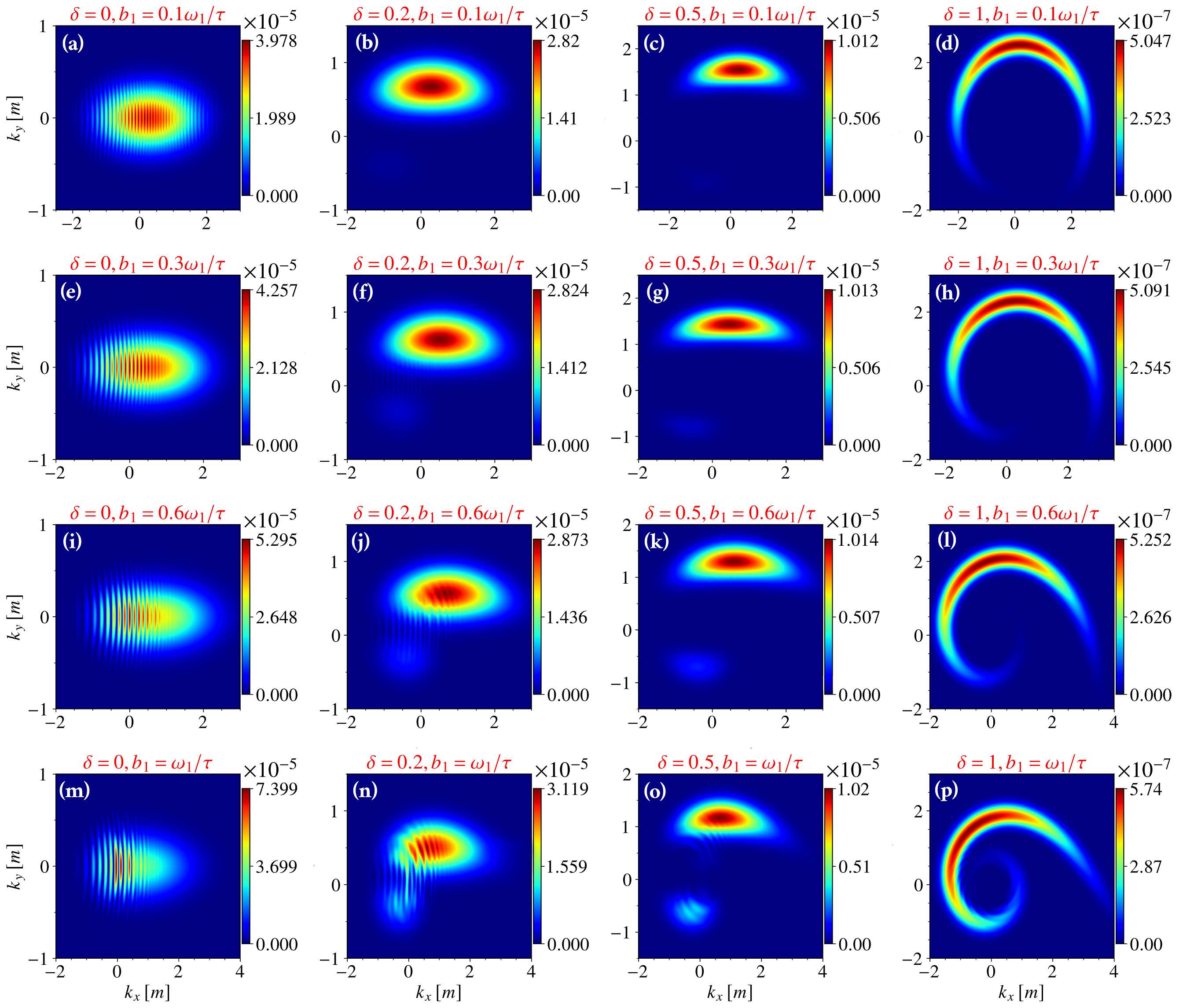}
		\caption{Momentum distribution of the created $e^+e^-$ pairs with $k_z = 0$ in presence of the one-color strong field $\mathbf{E}_{1s}$ with nonzero chirp $b_1$. The field parameters are given in Eq. \eqref{eq:parameters}.}
		\label{fig:MD_chirp_strong_strong_field}
	\end{figure*}
	The variation of the number density with polarization in case of two-color combined field $\mathbf{E}$, in Fig. \ref{fig:ND,EF_chirp_free_combined_field}(a), is not unique in general but depends on the chosen field parameters. In Ref. \cite{Olugh2020dynamically}, it was shown that variation of the number density with polarization is highly sensitive to the harmonic order of the weak field relative to the strong field i.e., the parameter $k$ introduced in Subsec. \ref{subsec:field_model}. Their results indicate that the number density increases with increasing harmonic order for each value of polarization, and that the dependence of the number density on the field polarization is nonlinear. Moreover, for lower harmonic orders, the number density decreases more rapidly as $\delta$ approaches $1$ compared with the case of higher harmonics. 
	
	To investigate the efficiency of the dynamical assistance in the two-color combined field $\mathbf{E}$, we plot in Fig. \ref{fig:ND,EF_chirp_free_combined_field}(b) the enhancement factor $\xi$, defined as \cite{aleksandrov2018dynamically, li2021enhanced, li2021study}
	\begin{equation}\label{eq:enhancement_factor_expression}
	 	\xi = \dfrac{N_{1s + 2w} }{N_{1s} + N_{2w}},
	\end{equation}
	as a function of the field polarization $\delta$ in the chirp-free case ($b_1 = b_2 = 0$). Here, $N_{1s + 2w}$, $N_{1s}$ and $N_{2w}$ are the number densities of the produced particles for the combined field, strong field and the weak field, respectively. 
	
	We observe a gradual increase in the enhancement factor from $\xi \sim 14$ at $\delta = 0$ to $\xi \sim 89$ at $|\delta| =  1$. This behavior arises because increasing $|\delta|$ suppresses the tunneling-driven particle yield of the strong field more efficiently that the multiphoton contribution of the weak field, thereby amplifying the relative role of dynamical assistance. Consequently, even though the absolute number density decrease with increase in $|\delta|$, the ratio defining the enhancement factor grows, indicating that the weak field becomes increasingly effective in assisting pair production at higher ellipticities. 

\section{Numerical results: Chirp applied to $\mathbf{E}_{1s}(t)$ only}\label{sec:results_chirp_strong}
	In this section, we present our numerical results for the one-color field configurations $\mathbf{E}_{1s}(t)$ and the two-color combined field $\mathbf{E}(t)$. In both the cases, we consider chirping only for $\mathbf{E}_{1s}(t)$, i.e., $b_1[\omega_1/\tau] = 0.1, 0.3, 0.6$ and $1$.
	\subsection{One-color strong field $\mathbf{E}_{1s}(t)$}\label{subsec:results_chirp_strong_strong_field}
	In this subsection, we investigate the momentum distribution and the number density of the produced $e^+e^-$ pairs in the one-color strong field $\mathbf{E}_{1s}$ with nonzero chirp $b_1$. Fig. \ref{fig:MD_chirp_strong_strong_field} shows the corresponding momentum distribution of the created pairs with variation in field polarization $\delta$ and the chirp parameter $b_1$.
	The variation of the overall distribution with field polarization $\delta$ is similar to the discussion made in Subsec. \ref{subsec:results_chirp_free_strong_field}. For linear polarization $\delta=0$, we observe the distribution to be symmetrical about $k_y$ axis, and elongated along the field direction with pronounced interference fringes, while increasing $\delta$ suppresses these fringes and shift the distribution toward finite $k_y$ region, eventually forming crescent-like structures in circular polarization limit ($\delta=1$). 
	
	On introducing frequency chirps, we observe the momentum distribution to be is highly sensitive to the chirp parameter $b_1$. As the chirp value is increased, the momentum distribution undergo a significant qualitative changes for all $\delta$. Chirping introduces time-dependent instantaneous frequency, which breaks the temporal symmetry of the pulse, hence we observe distorted and partial smeared interference fringes for $\delta=0$. This smeariness increases with increase in chirp value. For elliptically polarized fields $(\delta=0.2$ and $0.5)$, the combined effect of chirp and transverse field components enhances asymmetry in momentum space and promotes the formation of crescent-like structures, which are pronounced when field polarization is increased further to $\delta=1$. This reflects the interplay between the rotating electric field and the continuously changing frequency, which drives particles along curved trajectories in momentum space. 
	
	The peak value of the asymptotic momentum distribution function $f_\mathbf{k}(\infty)$ is increased gradually with increase in the chirp parameter $b_1$ for a fixed $\delta$. Meanwhile, for a fixed chirp strength, increasing $\delta$ reduces the peak value.
	
\begin{figure}[t]
		\centering
		\includegraphics[width=\linewidth]{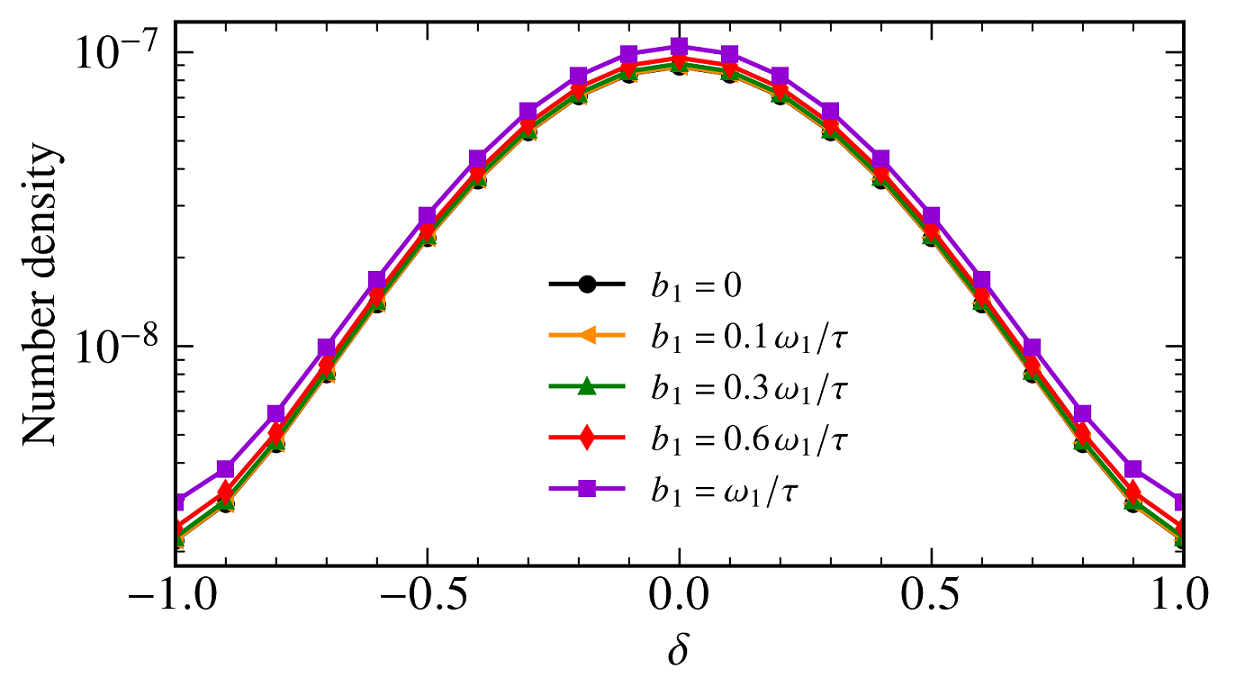}
		\caption{The number destiny (in units of $m^3$) of created particles as a function of field polarization $\delta$ for the one-color strong field $\mathbf{E}_{1s}$ with different chirp values $b_1$. The field parameters are given in Eq. \eqref{eq:parameters}.}
		\label{fig:ND_chirp_strong_strong_field}
	\end{figure}

	Fig. \ref{fig:ND_chirp_strong_strong_field} shows the number density of the created particles in presence of the one-color strong field $\mathbf{E}_{1s}$ as a function of field polarization $\delta$, and variation in chirp values of $b_1$. We observe that the number density exhibits a clear maxima at linear polarization $(\delta=0)$ and a monotonic suppression as $|\delta|$ is increased, for all values of $b_1$. We also observe that the number density shows only a weak dependence on the chirp parameter over the entire polarization range.
	Moreover, the highest number density is achieved for linearly polarized fields with the largest chirp, while elliptically and circularly polarized configurations produce significantly fewer pairs.

	\subsection{Two-color combined field $\mathbf{E}(t)$}\label{subsec:results_chirp_strong_combined_field}
	\begin{figure*}[tbh]
		\centering
		\includegraphics[width=\linewidth]{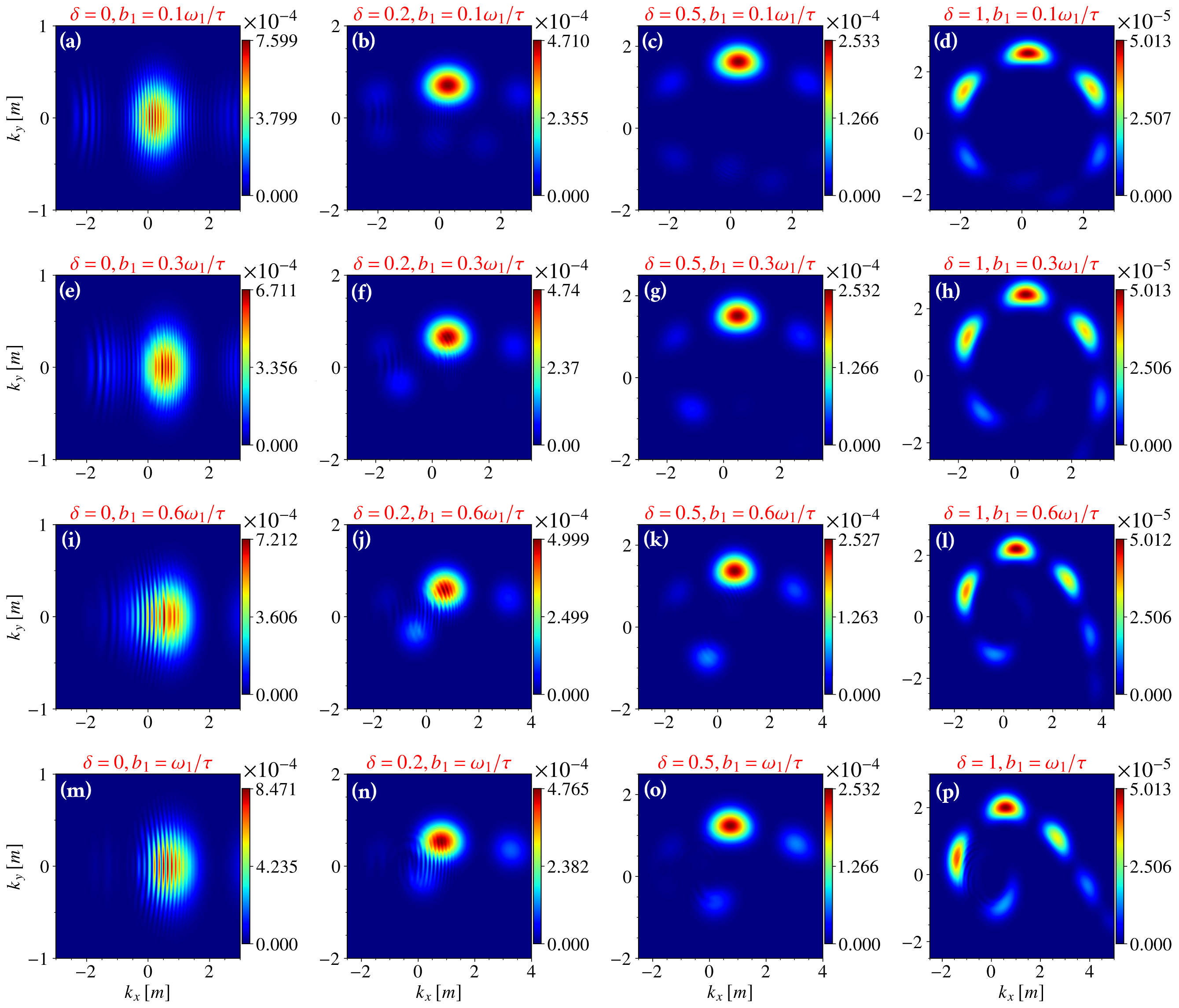}
		\caption{Momentum distribution of the created $e^+e^-$ pairs with $k_z = 0$ in presence of the two-color combined field $\mathbf{E}$ with chirp applied to $\mathbf{E}_{1s}$ only. The field parameters are given in Eq. \eqref{eq:parameters}.}
		\label{fig:MD_chirp_strong_combined_field}
	\end{figure*}
	In this subsection, we investigate the pair production in case of the two-color combined field $\mathbf{E}$ with chirp applied to $\mathbf{E}_{1s}$ only. 
	The momentum distribution of the created pairs with variation with field polarization $\delta$ and chirp values $b_1$ is shown in Fig. \ref{fig:MD_chirp_strong_combined_field}. 
	
	For the linear polarization $(\delta = 0)$, the distribution is symmetric about the $k_y=0$ axis, with pronounced interference fringes similar to $\delta = 0$ case in Fig. \ref{fig:MD_chirp_free_strong_field} and \ref{fig:MD_chirp_free_combined_field}. As the chirp value $b_1$ increases, these fringes exhibit smeariness along the $k_x$ direction, and the peak momentum distribution value rises, similar to $\delta = 0$ case in Fig. \ref{fig:MD_chirp_strong_strong_field}. For $\delta =0.2$ and $0.5$, the distribution shifts toward positive $k_y$ region, signaling the onset of the transverse momentum transfer due to the rotating field vector. In this case, the chirp mainly increases the peak momentum distribution value while preserving the overall spectrum structure. In the circularly polarized limit $\delta=1$, the momentum distribution forms a fragmented ring-like structure with peak momentum distribution values increasing gradually with the chirp value $b_1$. 
	
	Overall, increasing the polarization $\delta$ at a fixed $b_1$ value, continuously redistributes the momentum spectra from a narrow, interference-rich structure to a broadened fragmented ring-like structure while strongly suppressing the total pair yield. In contrast, increasing the chirp $b_1$, at a fixed value of $\delta$, gradually enhance the production rate and fringe visibility at small $\delta$($\sim 0$). 
	
	\begin{figure}[tbh]
	 	\centering
	 	\includegraphics[width=\linewidth]{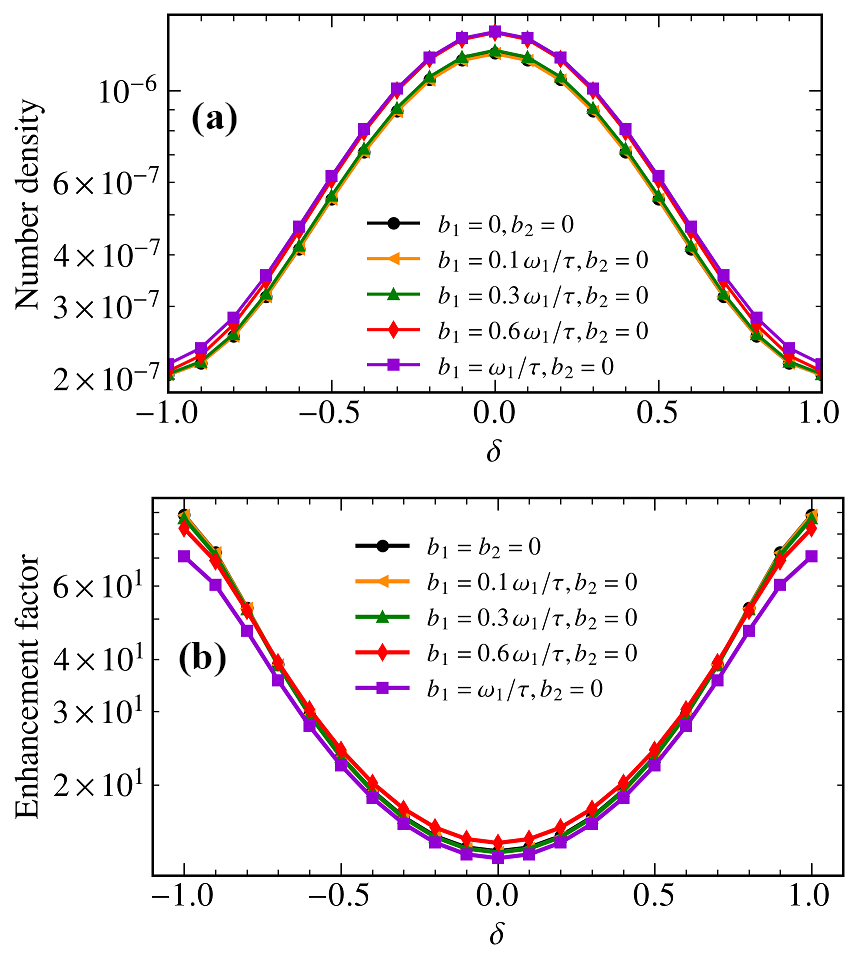}
	 	\caption{Chirp applied to the strong field $\mathbf{E}_{1s}$ only. 
	 		(a) The number density (in units of $m^{3}$) of created particles as a function of the field polarization $\delta$ for the two-color combined field $\mathbf{E}=\mathbf{E}_{1s}+\mathbf{E}_{2w}$. 
	 		(b) Corresponding enhancement factor of the total particle number in the combined field $\mathbf{E}$. 
	 		The field parameters are given in Eq.~\eqref{eq:parameters}.}
	 	\label{fig:ND,EF_chirp_strong_combined_field}
	\end{figure}
	In Fig. \ref{fig:ND,EF_chirp_strong_combined_field}(a), we plot the number density of the created particle with variation in field polarization for different chirp values $b_1$ in two-color combined field $\mathbf{E}(t)$. We observe that for all $b_1$, the number density reaches its maximum for linear polarization $\delta = 0$ and decrease monotonically toward elliptical and circular polarization $(|\delta|\rightarrow 1)$. Overall, the chirp parameter has only a weak influence on the number density across the entire polarization range, with the maxima being most noticeable at $\delta = 0$. In particular, chirp values $b_1[\omega_1/\tau] = 0.6$ and $1$ show a visible increase in the number density, while the curves corresponding to the other chirp values largely overlap. 

	Fig. \ref{fig:ND,EF_chirp_strong_combined_field}(b) shows the enhancement factor for the total number as a function of field polarization $\delta$ for different chirp values $b_1$ in two-color combined field $\mathbf{E}(t)$. We observe,  a minima at $\delta = 0$ (at a fixed $b_1$), and gradually increasing values toward large ellipticity. This is similar to the enhancement factor observed in Fig. \ref{fig:ND,EF_chirp_free_combined_field}(b) in the chirp-free case. The enhancement curves for different chirp strengths remain nearly identical, with only a slight decrease in the enhancement factor as the chirp increases. This suggests that the polarization dependence of the enhancement factor is largely insensitive to the chirp applied to the strong field, with chirp producing only a minor quantitative modification.

\section{Numerical results: Chirp applied to $\mathbf{E}_{2w}(t)$ only}\label{sec:results_chirp_weak}
	In this section, we present our numerical results for the one-color field $\mathbf{E}_{2w}(t)$ and the two-color combined field $\mathbf{E}(t)$. In both the cases, we consider chirping only for $\mathbf{E}_{2w}(t)$, i.e., $b_2[\omega_2/\tau] = 0.1, 0.3, 0.6$ and $1$.
	\subsection{One-color weak field $\mathbf{E}_{2w}(t)$}\label{subsec:results_chirp_weak_weak_field}
	\begin{figure*}[tbh]
		\centering
		\includegraphics[width=\linewidth]{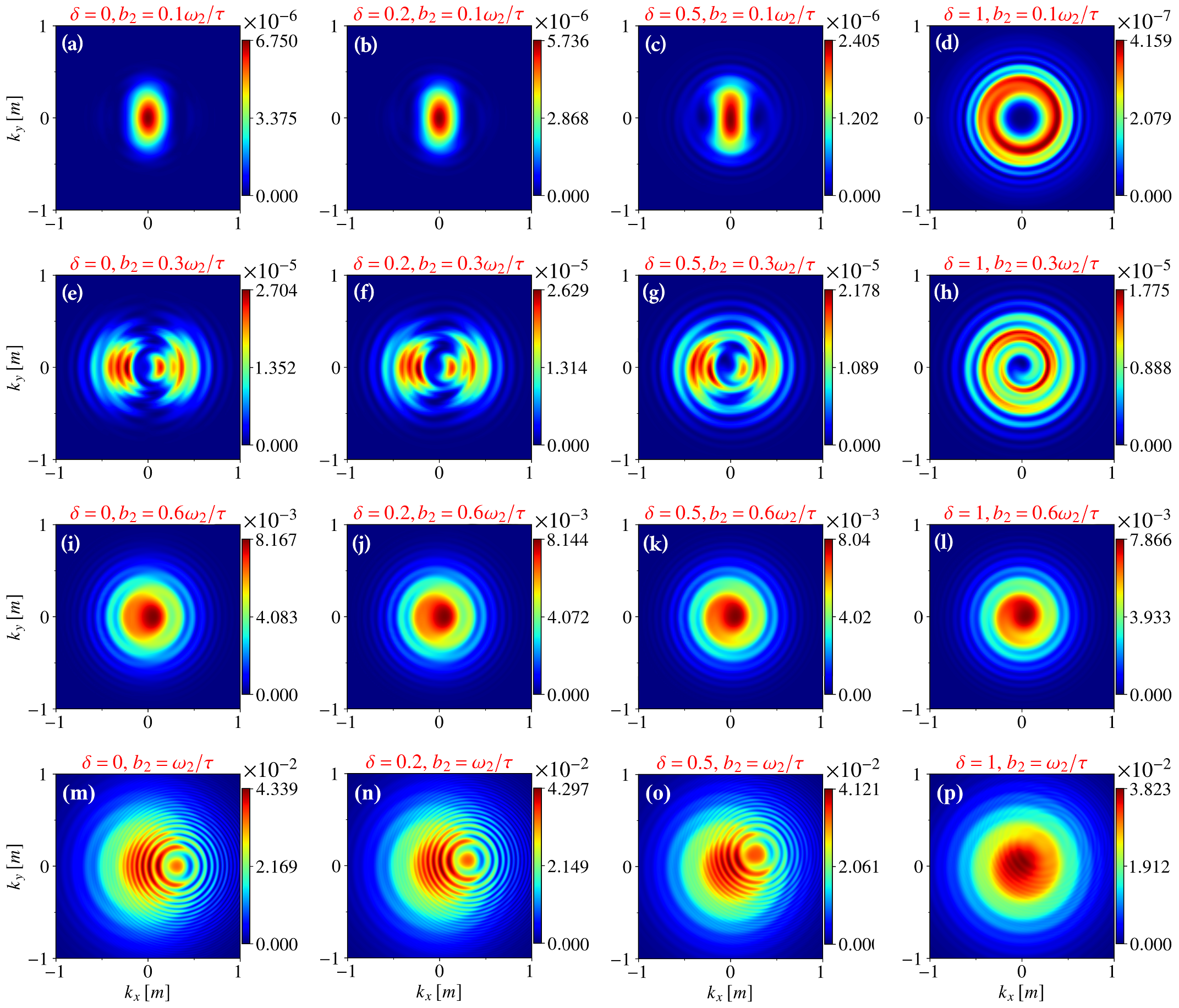}
		\caption{Momentum distribution of the created $e^+e^-$ pairs with $k_z = 0$ in presence of the one-color weak field $\mathbf{E}_{2w}$ with nonzero chirp $b_2$. The field parameters are given in Eq. \eqref{eq:parameters}.}
		\label{fig:MD_chirp_weak_weak_field}
	\end{figure*}
	Fig. \ref{fig:MD_chirp_weak_weak_field} shows the momentum distribution of the created pairs in one-color weak field $\mathbf{E}_{2w}$ with variation in field polarization $\delta$ and different chirp values $b_2$. For small chirp strength ($b_2[\omega_2/\tau] = 0.1 $), the momentum spectra remain strongly localized near the origin for linear polarization ($\delta=0$). The fragmented ring-like structure as seen in the chirp-free case, Fig. \ref{fig:MD_chirp_free_weak_field}(a), is no longer visible when chirp is applied. This is because: the ring-like structure observed for the chirp-free weak field originates from multiphoton resonance. Meanwhile, the introduction of even a small chirp broadens the resonance condition, causing the ring structure to collapse into a localized momentum distribution patch. 
	As $\delta$ increases, the distribution gradually broadens and develops weak circular features, signaling the onset of transverse momentum transfer induced by the rotating electric field component.
	At intermediate chirp values ($b_2[\omega_2/\tau] = 0.3$ and $0.6$), the spectra exhibit the spiral-like patterns. These pattern get pronounced with increase in chirp value and field polarization, reflecting the competition between chirp-driven temporal coherence and polarization driven rotational dynamics. 
	For the strongest chirp value $b_2[\omega_2/\tau] = 1$, the momentum distributions are dramatically enhanced with more number of crescent-like structure visible in the spectra structure. The spectra develop bright near-central accumulations surrounded by smooth circular-crescent patterns. 
	
	The peak value of the momentum distribution increases rapidly with increasing chirp strength. In contrast, increasing the polarization $\delta$ gradually reduces the peak height and redistributes strength over a wider momentum range, with chirp partially compensating this suppression at larger values. This variation of the peak value with field polarization is contrary to the case of one-color weak field without chirp, cf. Fig. \ref{fig:MD_chirp_free_weak_field}. We believe this is because: In the chirp-free case, increasing polarization enhances multiphoton absorption channels while particles still accumulate on the same resonance momentum ring, leading to a higher peak value of the distribution. In contrast, when chirp is present the resonance condition is smeared and polarization mainly redistributes particles over a wider transverse momentum region, hence reducing the peak value for a fixed chirp.
	
	\begin{figure}[tbh]
		\centering
		\includegraphics[width=\linewidth]{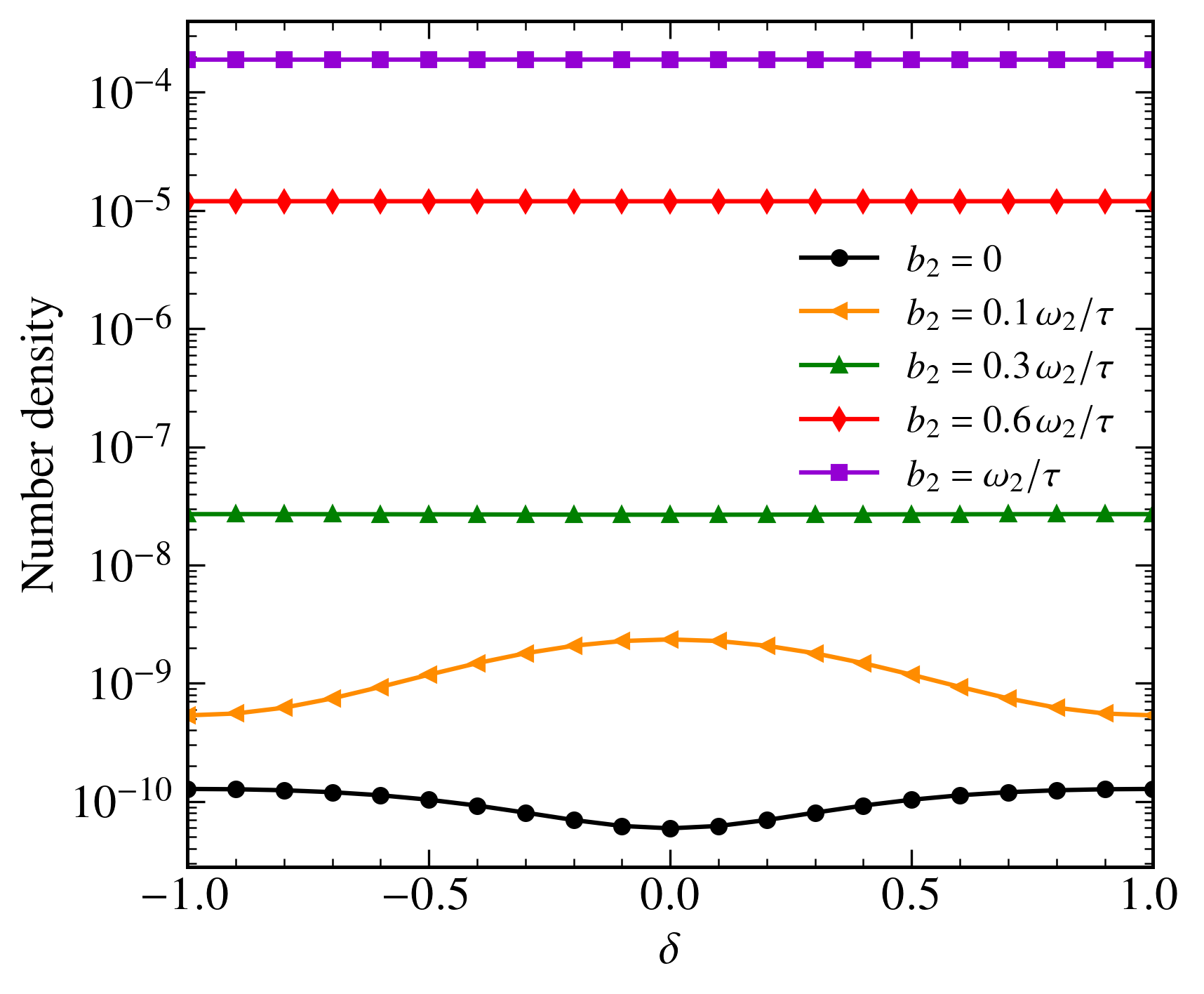}
		\caption{The number destiny (in units of $m^3$) of created particles as a function of field polarization $\delta$ for the one-color weak field $\mathbf{E}_{2w}$ with different chirp values $b_2$. The field parameters are given in Eq. \eqref{eq:parameters}.}
		\label{fig:ND_chirp_weak_weak_field}
	\end{figure}
	Fig. \ref{fig:ND_chirp_weak_weak_field} shows the number density as a function of field polarization in one-color weak field $\mathbf{E}_{2w}$ with different chirp values $b_2$. A small chirp value $b_2[\omega_2/\tau] = 0.1$ increases the yield by nearly an order of magnitude and leads to a pronounced maximum at $\delta=0$. With further increase in the chirp, the number density rises by several orders of magnitude and becomes almost insensitive to polarization. For the strongest chirp $b_2[\omega_2/\tau] = 1$, the yield reaches its maximum and remains essentially constant with respect to $\delta$, showing that strong chirping dominates over polarization effects. 
	
	Overall, these results demonstrate that in a one-color weak field $\mathbf{E}_{2w}$, chirp is the primary control parameter governing pair production, while the influence of polarization becomes negligible once the frequency modulation is sufficiently strong. 
	
	\subsection{Two-color combined field $\mathbf{E}(t)$}\label{subsec:resutls_chirp_weak_combined_field}
	In this subsection, we investigate the pair production in two-color combined field $\mathbf{E}(t)$ with chirp applied to $\mathbf{E}_{2w}$ only. 
	\begin{figure*}[tbh]
		\centering
		\includegraphics[width=\linewidth]{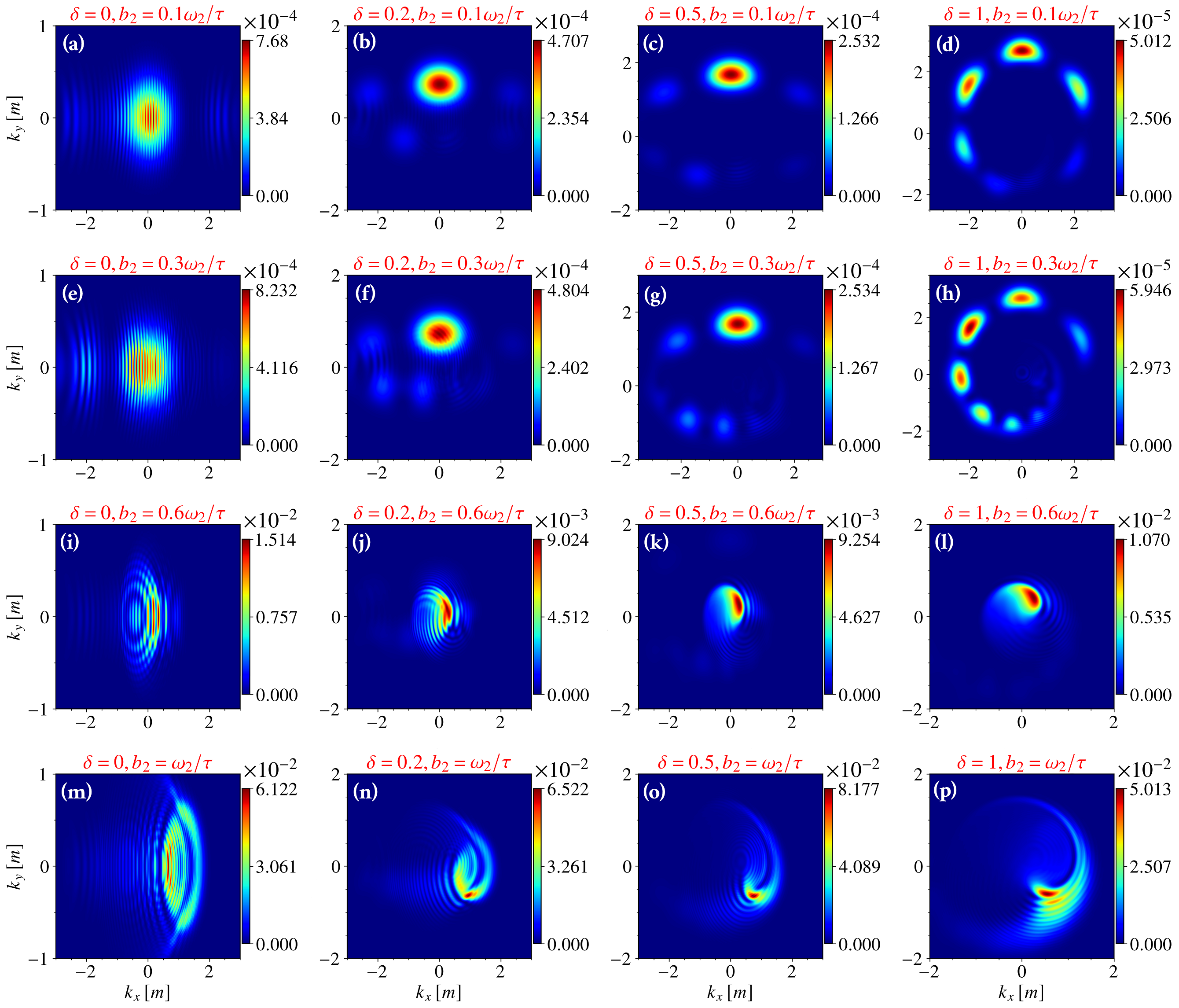}
		\caption{Momentum distribution of the created $e^+e^-$ pairs with $k_z=0$ in two-color combined field $\mathbf{E}$ with chirp applied to $\mathbf{E}_{2w}$ only. The field parameters are given in Eq. \eqref{eq:parameters}.}
		\label{fig:MD_chirp_weak_combined_field}
	\end{figure*}
	Fig. \ref{fig:MD_chirp_weak_combined_field} shows the corresponding momentum distribution of the created pairs with different $\delta$ and chirp values $b_2$. 
	
	For the linear polarization $\delta=0$, we observe the spectra to be dominated by a single, centrally localized peak elongated on the $k_x$ direction, reflecting the dominance of the strong field component and the effective one-dimensional acceleration of the created pairs. As the chirp parameter $b_2$ is increased, the peak magnitude rises significantly. 
	With increasing polarization ellipticity $(\delta=0.2$ and $0.5)$, the momentum spectra shifts towards the positive $k_y$ region and evolve into asymmetric structures which accounts for the growing transverse momentum transfer caused by the rotating field vector. The peak values generally decrease compared to the linear case.  
	In the circular polarization $\delta=1$ with small chirp $b_2[\omega_2/\tau] = 0.1$ and $0.3$, the momentum spectra form fragmented ring like structure which turns into a more complex spiral-like structure when the chirp $b_2$ is increased further gradually. 
	
	Overall, the results demonstrate that polarization primarily controls the global structure and symmetry of the momentum distribution, while chirp acts as an efficient control knob for enhancing peak values. Here, the chirp is applied only to the weak field, its role is not to dominate the dynamics but to resonantly assist the strong field by modulating the temporal structure of the combined field. This leads to enhanced pair creation and richer momentum-space features, particularly for linear and elliptic polarizations, highlighting the potential of weak-field chirping as a powerful tool for coherent control of strong-field vacuum pair production.
	
	In Fig. \ref{fig:ND,EF_chirp_weak_combined_field}(a), we plot the number density as a function of the field polarization for the two-color combined field $\mathbf{E}$, where the chirp is applied only to the weak field component $\mathbf{E}_{2w}$. We observe a clear dependence on both the polarization parameter $\delta$ and the chirp parameter $b_2$. For small chirp values, i.e., $b_2[\omega_2/\tau] = 0.1$ and $0.3$, the behavior of the number density with respect to $\delta$ is quite similar: it peaks around linear polarization $(\delta = 0)$ and decreases toward circular polarization $(|\delta| = 1)$. 
	
	As the chirp parameter $b_2$ increases, the number density gradually increases. However, the sensitivity to the field polarization is suppressed, as the curves become flatter and the number density becomes nearly independent of $\delta$. Physically, strong chirping broadens the frequency spectrum and introduces rapid phase variations, which reduce the sensitivity to the geometric structure of the field polarization. In other words, the chirp-driven enhancement dominates over the polarization-dependent suppression. We note the enhancement in number density up to 2$-$3 orders when the strongest chirp, i.e., $b_2[\omega_2/\tau] = 1$ is applied in such case. 
	\begin{figure}[tbh]
		\centering
		\includegraphics[width=\linewidth]{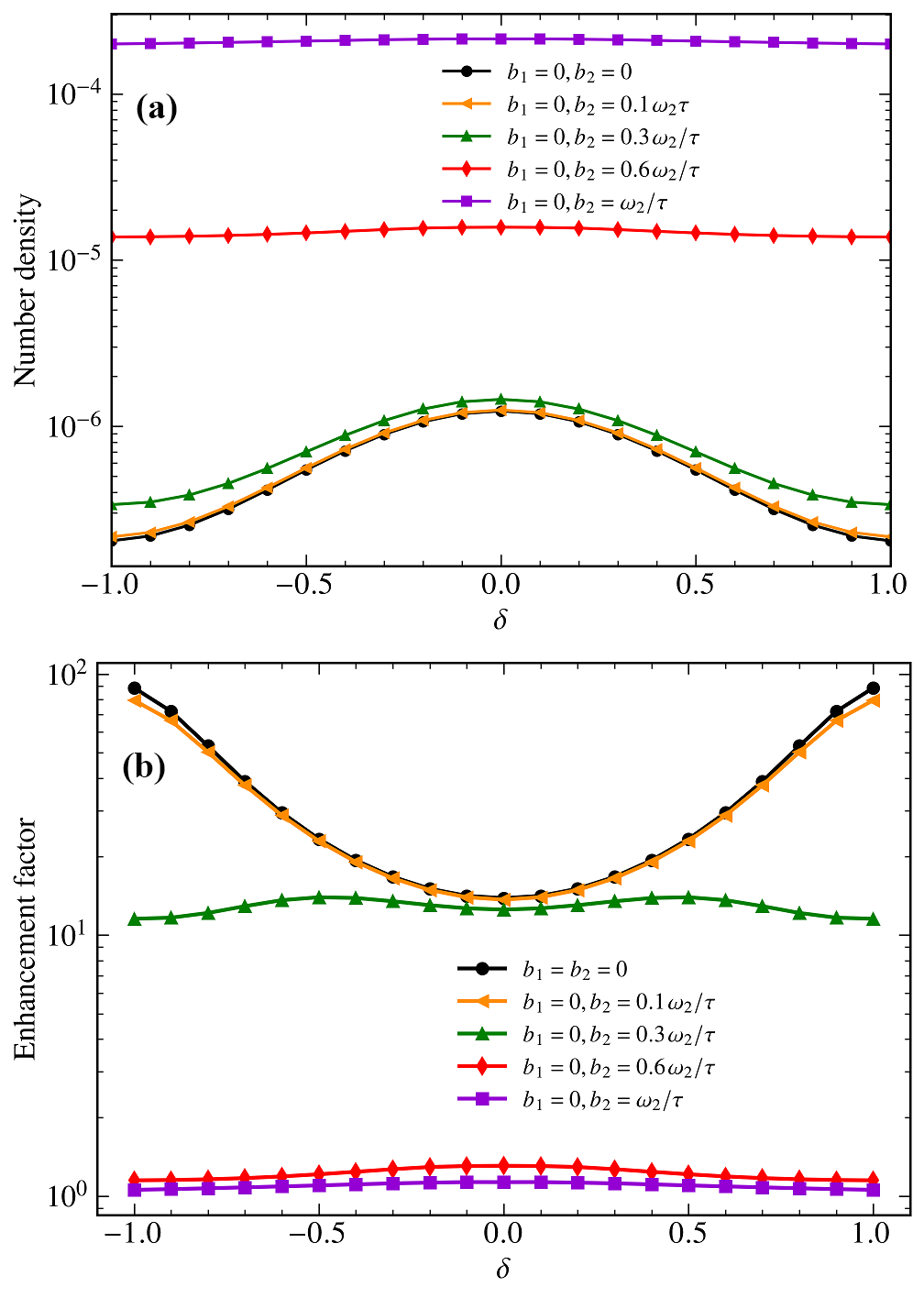}
		\caption{Chirp applied to the weak field $\mathbf{E}_{2w}$ only. 
			(a) The number density (in units of $m^{3}$) of created particles as a function of the field polarization $\delta$ for the two-color combined field $\mathbf{E}=\mathbf{E}_{1s}+\mathbf{E}_{2w}$. 
			(b) Corresponding enhancement factor of the total particle number in the combined field $\mathbf{E}$. 
			The field parameters are given in Eq.~\eqref{eq:parameters}.}
		\label{fig:ND,EF_chirp_weak_combined_field}
	\end{figure}
	
	In Fig.~\ref{fig:ND,EF_chirp_weak_combined_field}(b), we plot the enhancement factor for the two-color combined field $\mathbf{E}$, where the chirp is applied only to the weak field component $\mathbf{E}_{2w}$, as a function of the field polarization $\delta$ for different chirp values $b_2$. For the unchirped case and for a small chirp ($b_2=0.1\omega_2/\tau$), the enhancement factor exhibits a minimum at linear polarization ($\delta=0$) and increases significantly toward circular polarization ($|\delta|\rightarrow1$), indicating that the relative gain due to dynamical assistance is greatest for circularly polarized fields. As the weak-field chirp is increased, however, both the magnitude of the enhancement factor and its dependence on polarization are progressively suppressed. For $b_2=0.3\omega_2/\tau$, the enhancement factor develops a distinct wing-like profile, exhibiting minima at $|\delta|=1$ and broad maxima at intermediate ellipticities., while for $b_2=0.6\omega_2/\tau$ and $b_2=\omega_2/\tau$, it remains close to unity over the entire polarization range. This demonstrates that increasing the chirp of the weak field diminishes the relative contribution of dynamical assistance, even though the absolute pair yield continues to increase, because the chirped weak field itself becomes increasingly effective at producing pairs. Consequently, the additional enhancement arising from the combined-field configuration is substantially reduced.

\section{Numerical results: Chirp applied to both $\mathbf{E}_{1s}(t)$ and $\mathbf{E}_{2w}(t)$}\label{sec:results_chirp_strong_weak}
	In this section, we investigate the pair production in presence of two-color combined field $\mathbf{E} = \mathbf{E}_{1s} + \mathbf{E}_{2w}$ with chirp applied to both strong field $\mathbf{E}_{1s}$ and weak field $\mathbf{E}_{2w}$, simultaneously. For simplicity and meaningful comparison of the results, we adopt a rule for selecting the chirp parameters $b_1$ and $b_2$ as $b_1/b_2 = \omega_1/\omega_2$, such that both the strong and weak field components are chirped with equal relative strength i.e., $b_1 = a \omega_1/\tau$ and $b_2 = a\omega_2/\tau$ where $a$ is a dimensionless control parameter. We vary $a$ as $0.1, 0.3, 0.6$ and $1$.
	\begin{figure*}[t]
		\centering
		\includegraphics[width=\linewidth]{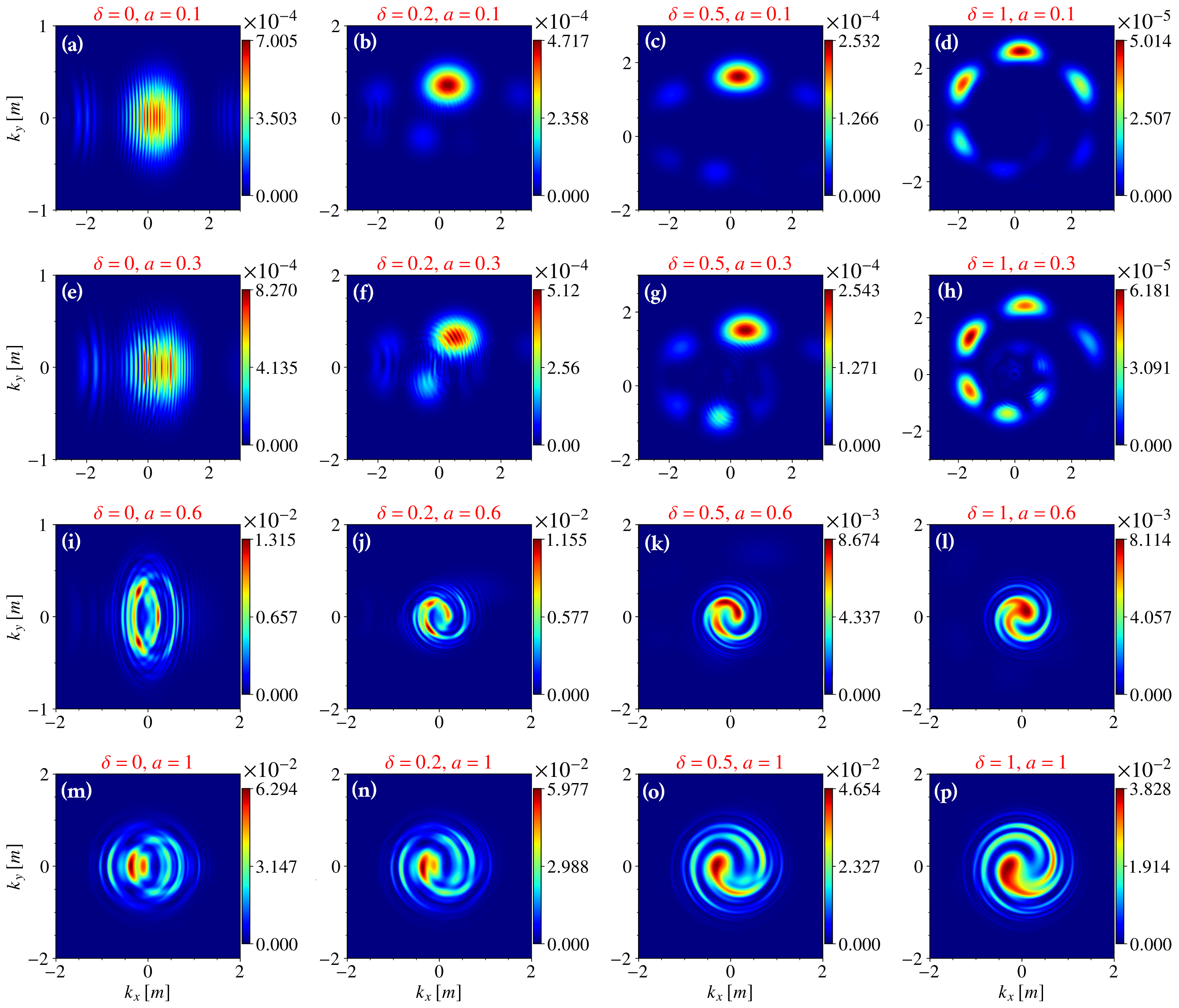}
		\caption{Momentum distribution of the created $e^+e^-$ pairs with $k_z=0$ in two-color combined field $\mathbf{E}$ with chirp applied to both $\mathbf{E}_{1s}$ and $\mathbf{E}_{2w}$. The chirp parameters $b_1$ and $b_2$ are varied together through the dimensionless control parameter $a$, defined as $b_1 = a\omega_1/\tau$ and $b_2 = a\omega_2/\tau$. The other field parameters are given in Eq. \eqref{eq:parameters}.}
		\label{fig:MD_chirp_strong_weak_combined_field}
	\end{figure*}
	In Fig. \ref{fig:MD_chirp_strong_weak_combined_field}, we plot the momentum distribution function for the created $e^+e^-$ pairs for different $\delta$ and $a$.
	
	For small chirp $(a = 0.1)$, the distribution retains features characteristic of polarization. At $\delta=0$, the spectrum is elongated along the field direction with pronounced interference fringes. As $\delta$ increases to $0.2$ and $0.5$, the distribution becomes more localized in the transverse momentum plane and develops multiple separated lobes. In the circular limit ($\delta=1$), the spectrum forms a fragmented ring like structure, reflecting multiphoton channel interference. As the chirp strength increase ($a = 0.3$ and $0.6$), the distributions undergo clear rotational distortion. The lobes bend and gradually transform into spiral-like patterns, especially for $\delta\geq0.2$. The spiral structure becomes very pronounced at $a=1$, where all polarization cases exhibit a rotating, vortex-like momentum distribution. 
	
	The overall trend of the peak value is the gradual increment with increase in the chirp control parameter $a$ while gradual decrement with increase in the field polarization $\delta$. This is because of the suppression of tunneling in more circular polarization. As a result, peak value decreases as $\delta$ increases. While the increment of the peak value with increase in the chirp is due to the improvement of the temporal overlap of the field. The parameter $a$ primarily governs the strength of enhancement and the transition from interference-dominated patterns to spiral/vortex structures, whereas $\delta$ controls the geometric symmetry and angular redistribution of momentum. The combined action of frequency chirp in both fields and polarization rotation leads to strong reshaping and significant quantitative enhancement of the peak momentum density. 
	
	\begin{figure*}[tbh]
	\centering
	\includegraphics[width=\linewidth]{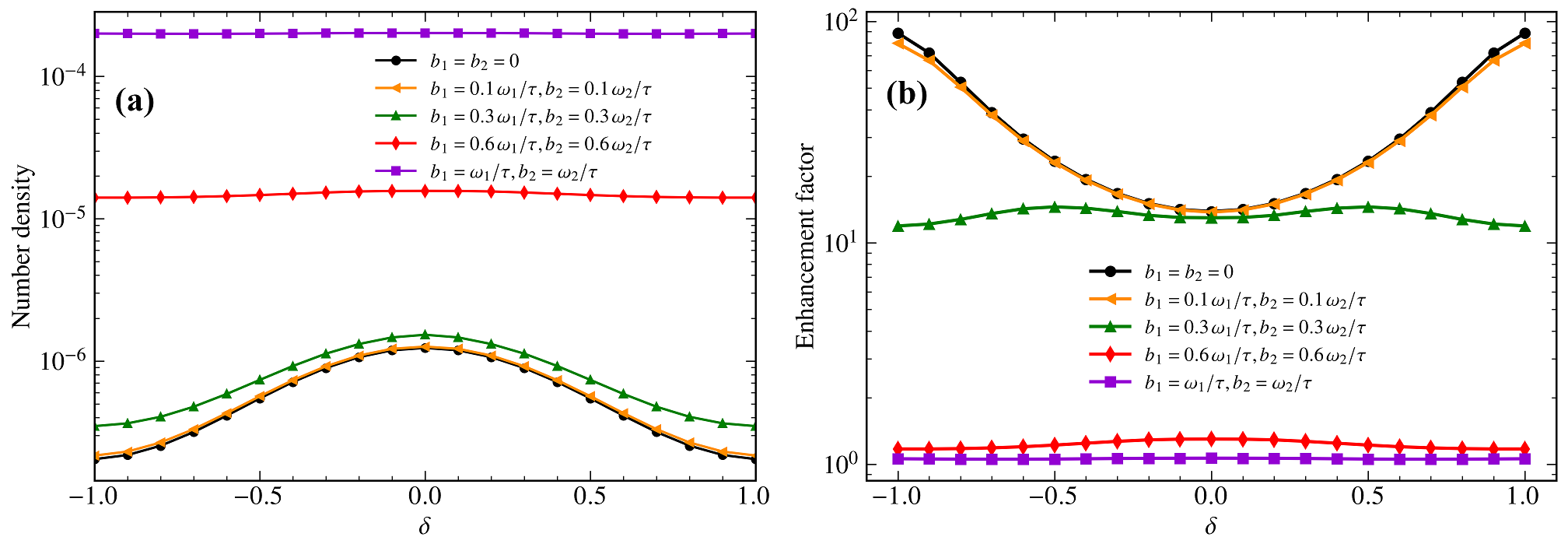}
	\caption{Chirp applied to both $\mathbf{E}_{1s}$ and $\mathbf{E}_{2w}$. 
		(a) The number density (in units of $m^{3}$) of created particles as a function of the field polarization $\delta$ for the two-color combined field $\mathbf{E}=\mathbf{E}_{1s}+\mathbf{E}_{2w}$. 
		(b) Corresponding enhancement factor of the total particle number in the combined field $\mathbf{E}$. 
		The field parameters are given in Eq.~\eqref{eq:parameters}.}
	\label{fig:ND,EF_chirp_strong_weak_combined_field}
	\end{figure*}
	In Fig. \ref{fig:ND,EF_chirp_strong_weak_combined_field}(a), we plot the number density as a function of field polarization for the two-color combined field $\mathbf{E}(t)$ for different chirp parameters $b_1$ and $b_2$. The overall behavior closely resembles the case where only the weak field is chirped, cf. Fig. \ref{fig:ND,EF_chirp_weak_combined_field}(a). For small chirps ($b_1[\omega_1/\tau]=b_2[\omega_2/\tau]=0.1, 0.3$), the number density is maximum near linear polarization ($\delta=0$) and decreases toward circular polarization ($|\delta|=1$). Increasing the chirp enhances the absolute pair yield while progressively reducing its dependence on polarization, so that for $b_1[\omega_1/\tau]=b_2[\omega_2/\tau] = 1$ the number density becomes nearly independent of $\delta$. Thus, simultaneous chirping of both fields leads to the same qualitative behavior as chirping only the weak field, with a comparable enhancement of the pair yield.
	
	Figure~\ref{fig:ND,EF_chirp_strong_weak_combined_field}(b) presents the corresponding enhancement factor. Similar to the previous case i.e., Fig. \ref{fig:ND,EF_chirp_weak_combined_field}(b), the unchirped and weakly chirped fields exhibit a minimum at $\delta=0$ and increase toward circular polarization. For $b_1[\omega_1/\tau]=b_2[\omega_2/\tau] = 0.3$, the enhancement factor develops the same characteristic wing-like profile with broad maxima at intermediate ellipticities. As the chirp strength is increased further, the enhancement factor approaches unity and becomes only weakly dependent on polarization, indicating that although simultaneous chirping substantially increases the absolute pair yield, it reduces the relative contribution of dynamical assistance. The physical mechanism behind this result/phenomenon needs to be studied further.
	
	Thus, chirp acts as a powerful control parameter that enhances the overall yield while simultaneously suppressing polarization dependence in the two-color combined field. Also, strong chirping of both fields suppresses the nonlinear cooperation between them, making the pair production more additive rather than dynamically assisted.

\section{Conclusion and Discussion}\label{sec:Conclusion}
	Within the real-time DHW formalism, we investigated the combined effects of linear frequency chirps and field polarization on electron-positron ($e^+e^-$) pair production in three field configurations: a one-color strong-slowly varying field, a one-color weak-rapidly varying field, and their dynamically assisted two-color combinational field. Our numerical results are presented in terms of the number density and momentum distribution of the produced pairs, as well as the enhancement factor in the dynamically assisted case. In particular, we analyzed pair production for these field configurations under different chirp conditions : (i) chirp-free fields, (ii) chirp applied only to the strong field $\mathbf{E}_{1s}$, (iii) chirp applied only to the weak field $\mathbf{E}_{2w}$, and (iv) chirp applied to both strong and weak fields, simultaneously. The main findings of this study are summarized as follows.
	
	In the chirp-free case, we examined one-color strong and weak fields as well as their two-color dynamically assisted combination. In all three cases, we find the momentum distribution to be is highly sensitive to the field polarization. For the strong field, the momentum distribution exhibits elongated spectra with pronounced interference fringes in linear polarization, while increasing ellipticity suppress the interference and eventually leads to crescent-like structures for circular polarization. Correspondingly, the peak momentum distribution value and the number density decrease monotonically with increasing ellipticity due to the reduction of the effective tunneling component along a fixed direction. In contrast, for the weak field, the momentum spectra exhibit characteristic multiphoton ring structures whose peak values and corresponding number density increase with polarization. When both fields are combined, the dynamically assisted Schwinger mechanism enhances the particle yield significantly compared with either field alone, while the polarization dependence remains qualitatively similar to the tunneling-dominated strong-field behavior. 
	
	When chirp is introduced in the strong field alone, the momentum spectra become distorted and interference fringes gradually smear due to the time-dependent instantaneous frequency. However, the overall pair yield exhibits only moderate enhancement and the polarization dependence remains largely unchanged, indicating that the tunneling dynamics continue to dominate the production process. In contrast, chirping the weak field leads to a much stronger modification of the particle spectra and production rate. The momentum distributions develop spiral- and crescent-like structures as the chirp increases, reflecting the interplay between the rotating electric field and the time-dependent frequency modulation. More importantly, the number density increases by several orders of magnitude for sufficiently strong chirp. In this regime the pair yield becomes almost insensitive to the field polarization, indicating that chirp-induced nonadiabatic frequency sweeping becomes the dominant mechanism controlling pair production.
	
	Finally, we investigated the case where both the strong and weak fields are chirped simultaneously. In this configuration, the momentum distributions evolve from interference-dominated patterns to vortex-like or spiral structures as the chirp strength increases. The particle yield is significantly enhanced compared with the chirp-free case, while the dependence on field polarization becomes progressively weaker. This demonstrates that strong frequency modulation can effectively override the geometric effects associated with field polarization. At the same time, we observe that the enhancement factor of the dynamically assisted mechanism may decrease for very strong chirp, suggesting that excessive chirping can reduce the nonlinear cooperation between the two fields and make the production process more additive rather than dynamically assisted.
	
	We find that the enhancement factor consistently reaches its maximum at circular polarization, $|\delta| = 1$, for all chirp configurations considered. Interestingly, we also find that among the different chirp cases, the enhancement factor is the largest, $\xi \sim 89$, in the chirp free case, $b_1= b_2 = 0$. 
	
	Overall, our results demonstrate that frequency chirp provides a powerful control parameter for manipulating vacuum pair production in strong laser fields. While polarization primarily determines the geometric structure of the momentum spectra and influences tunneling efficiency, chirp can strongly enhance particle production and reshape the spectral features through time-dependent frequency modulation. Chirping the fields substantially enhances the pair-production yield, with the enhancement becoming more pronounced as the chirp strength increases. A notable result of this work is that chirping only the weak field produces nearly the same increase in the total number density as chirping both the strong and weak fields simultaneously, indicating that the weak-field chirp plays the dominant role in the chirp-induced enhancement of the dynamically assisted Schwinger mechanism. In contrast, chirping only the strong field results in a comparatively modest increase in the pair yield. Furthermore, although the absolute number density increases with increasing chirp, the enhancement factor exhibits a distinctly different behavior. For the chirp-free and weakly chirped cases, the enhancement factor is largest for circular polarization ($|\delta|=1$), reflecting the stronger relative contribution of dynamical assistance at higher ellipticities. However, when the chirp is applied to the weak field alone or to both fields simultaneously, the enhancement factor decreases rapidly with increasing chirp and eventually approaches unity over the entire polarization range, indicating that the chirped fields themselves become increasingly efficient in producing pairs and hence reduce the relative gain from dynamical assistance.
	
	Finally, the present results indicate that the polarization dependence of the enhancement factor is governed by a nontrivial interplay between chirping and the dynamical-assistance mechanism. A more complete understanding of this behavior will require systematic parameter scans including independent variation of the pulse durations of both strong and weak field ($\tau_1$ and $\tau_2$), together with the high harmonic order $k$ ($ = \omega_2/\omega_1$), chirp strengths ($b_1$ and $b_2$), and polarization $\delta$. Such an extended analysis would help identify the parameter regimes in which dynamical assistance is optimized and clarify the transition between tunneling- and multiphoton-dominated dynamics. This investigation lies beyond the scope of the present work and is reserved for future studies.
	
	Our work provides useful insights into the coherent control of nonperturbative pair production processes and may serve as a guide for optimizing laser-field configurations in future high-intensity laser facilities aimed at probing strong-field quantum electrodynamics.

%======================bibliography================
\end{document}